# Shock-Sensitivity in Shell-Like Structures: with Simulations of Spherical Shell Buckling

J. Michael T. Thompson
*Department of Applied Maths and Theoretical Physics,*
*University of Cambridge, CB3 0WA, UK*
and *Department of Engineering, Aberdeen University*

Jan Sieber
*CEMPS, University of Exeter, Exeter EX4 4QF, UK*

Under increasing compression, an unbuckled shell is in a metastable state which becomes increasingly precarious as the buckling load is approached. So to induce premature buckling a lateral disturbance will have to overcome a decreasing energy barrier which reaches zero at buckling. Two archetypal problems that exhibit a severe form of this behaviour are the axially-compressed cylindrical shell and the externally pressurized spherical shell. Focussing on the cylinder, a non-destructive technique was recently proposed to estimate the 'shock sensitivity' of a laboratory specimen using a lateral probe to measure the nonlinear load-deflection characteristic. If a symmetry-breaking bifurcation is encountered on the path, computer simulations showed how this can be supressed by a controlled secondary probe. Here, we extend our understanding by assessing in general terms how a single control can capture remote saddle solutions: in particular how a symmetric probe could locate an asymmetric solution. Then, more specifically, we analyse the spherical shell with point and ring probes, to test the procedure under challenging conditions to assess its range of applicability. Rather than a bifurcation, the spherical shell offers the challenge of a de-stabilizing fold (limit point) under the rigid control of the probe.

## 1. Introduction

We have recently proposed a new non-destructive experimental procedure for determining the shock sensitivity of a shell-like structure in its 'trivial' metastable state below its buckling load [Thompson, 2015]. Simulations have shown how this works successfully for a cylindrical shell under axial compression, and here we extend this demonstration to a complete spherical shell under uniform external pressure. In this technique we *rigidly* apply a simple lateral load (such as one or more point forces), to determine a lateral-force, $Q$, versus lateral-displacement, $q$, graph. Continued until the lateral force has dropped to zero, we have then located a true equilibrium state of the free compressed shell, which in the absence of the lateral constraint would be unstable. The area under the $Q(q)$ force-displacement graph then supplies the energy barrier that is needed to surpass this free unstable saddle-point, H, in the total potential energy of the undisturbed structure. It can be noted that the severe sensitivity to geometric imperfections known to be present in many compressed shells means that any non-destructive experimental technique which explores the post-buckling behaviour is potentially useful.

A possible complication arises if there is a bifurcation in the $Q(q)$ curve, and we have shown how this can be overcome by the introduction of a second rigidly-controlled probe which is tuned to offer zero force.
A question naturally arises as to how successful a single probe can be in detecting a saddle with a more complex shape. Here, we first study this in general terms and show how mountain-pass saddles effectively 'attract' loading paths to themselves. We then analyse the buckling of a spherical shell into an axisymmetric mode. As we would expect, a point load on the shell easily locates the unstable axisymmetric dimpled form of saddle point H. We then use a distributed ring loading to test the limitations of the method when a wide ring is pressing inwards at a radius where the shell would naturally be deflecting outwards. We should note here that in our axisymmetric simulation the technique can naturally only locate the axisymmetric mode, but applied to a real shell in a test we show that it could equally well locate a non-symmetric mode, possibly in other problems finding the lowest of various energy passes.

## 2. The probing technique

The well-known potential energy topologies at the two basic pitchfork bifurcations [Thompson & Hunt, 1973] are illustrated in figure 1.

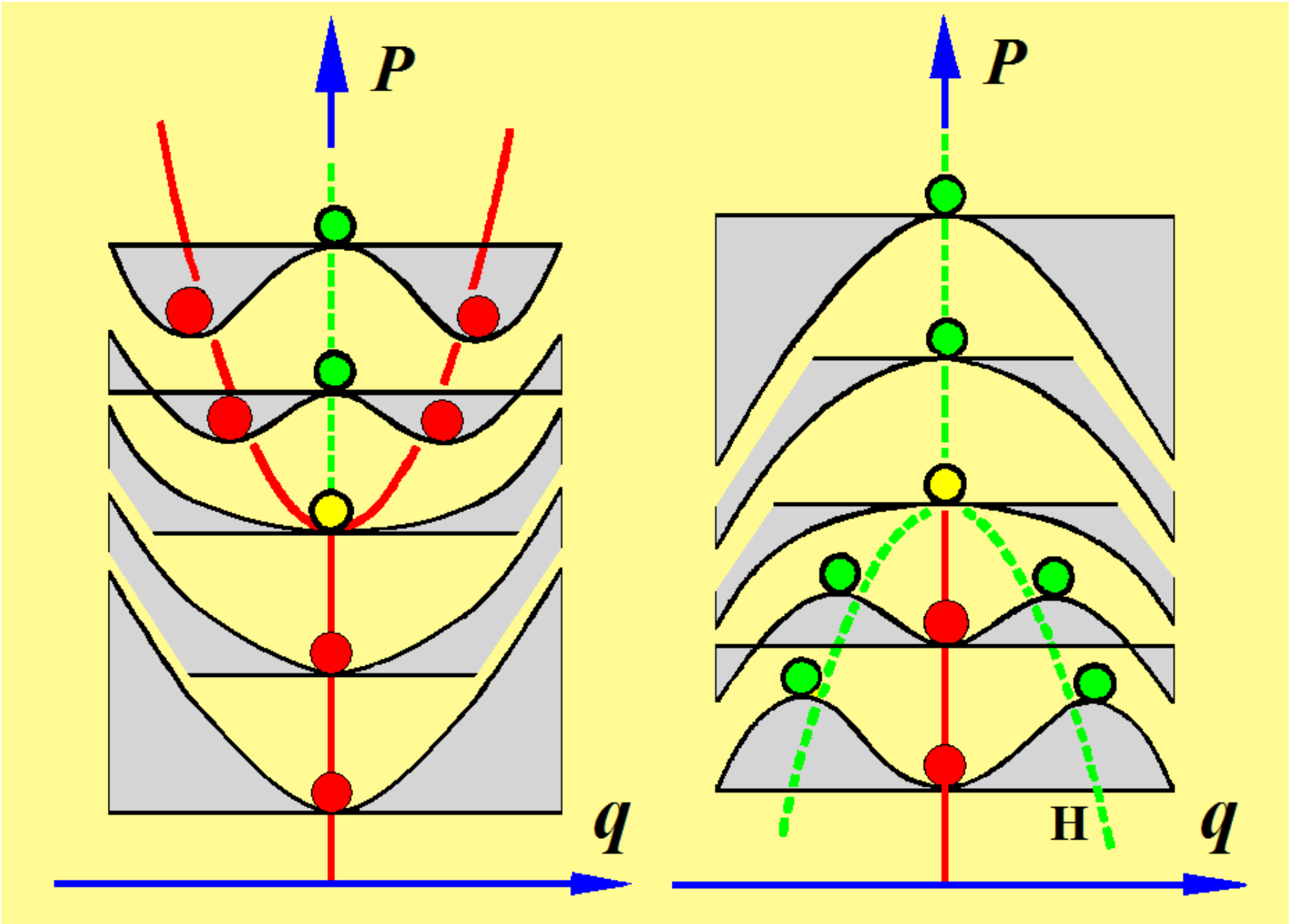

Fig 1. The super-critical and sub-critical pitchfork bifurcations on a plot of the load, *P*, against a generalised deflection, *q*. The sub-critical form, on the right-hand side, displays the metastability that is the topic of this article

The first picture shows the super-critical pitchfork with a rising and stable post-buckling path which is familiar in the response of a compressed Euler column. The second picture shows the sub-critical pitchfork with a falling unstable post-buckling path similar to those encountered in the nonlinear post-buckling behaviour of thin compressed shell-like structures. In this second case, the trivial unbuckled state at a load, *P*, less than its critical value is *metastable*, that is to say it is stable against small disturbances but unstable against larger disturbances. In this regime, the probability of (premature) failure due to finite static or dynamic disturbances is governed by the height of the potential energy barrier located at the unstable post-buckling state, H. It is the experimental determination of this barrier that concerns us in this paper, because a low barrier implies that the structure will exhibit 'shock-sensitivity' [Thompson & van der Heijden, 2014].

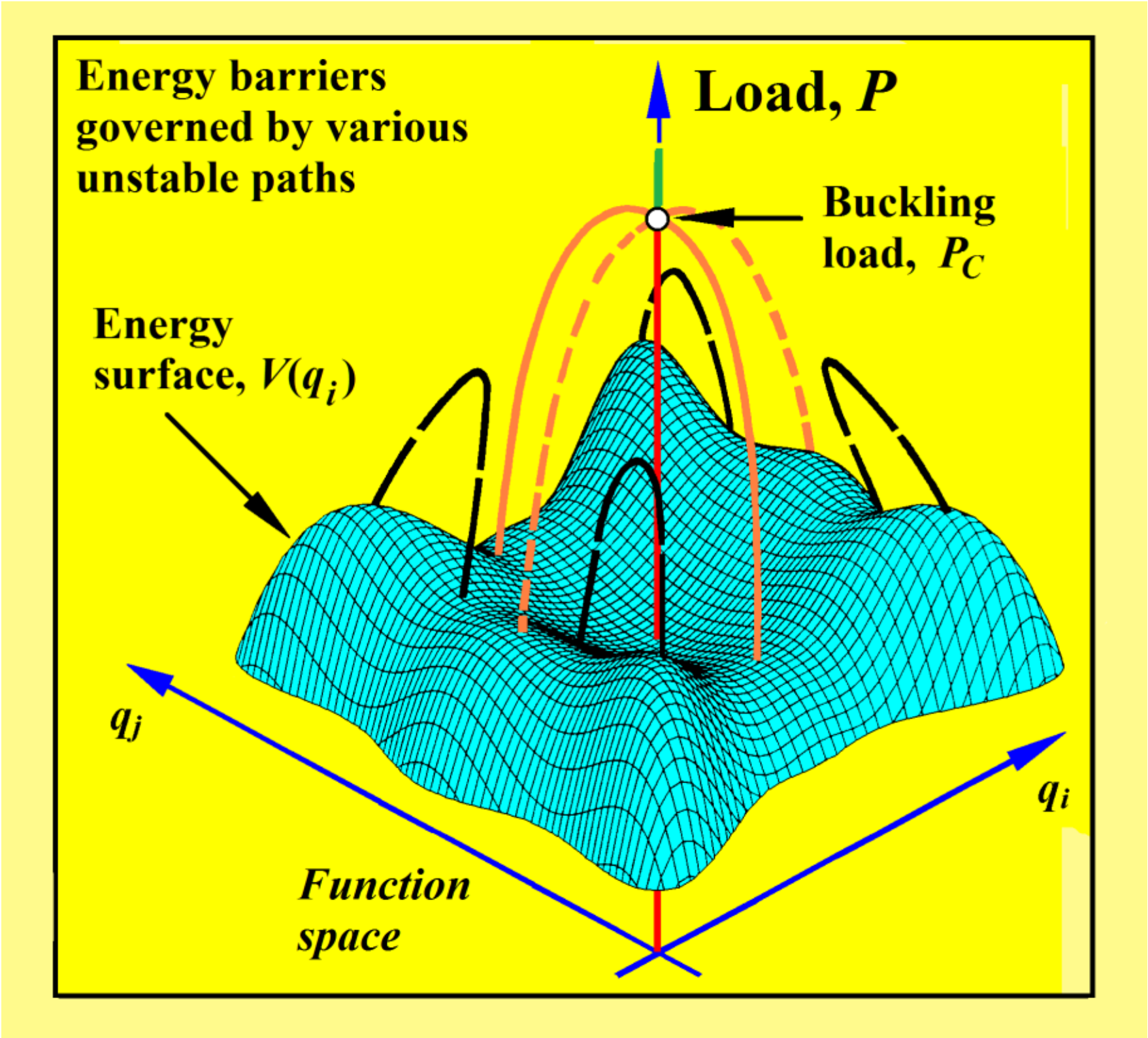

Fig 2. Alternative escape routes in higher dimensions. A schematic diagram showing notional post-buckling paths on a plot of load, *P*, against two generalized coordinates. At a load less than critical, there is superimposed a sketch of the total potential energy function $V(q_i)$.

Of course the situation illustrated in figure 1 for the two pitchforks is a gross simplification of the escape story for more complex structures, and a more realistic picture is sketched in figure 2 for a hypothetical shell-like structure with many post-buckling paths. This is still highly schematic, because the function space of the deformations is now represented by just two coordinates representing what in practice might be a large but finite set of generalised coordinates for a discrete (or discretized) system or the genuinely infinite-dimensional function space of an elastic continuum system. A few notional post-buckling paths are sketched, to represent the large number of such paths that are known to arise in long thin structures, and some are associated with 'mountain passes' over which the system could escape: others can be associated with local maxima, as can be seen in the figure. Note that rather than emerging from a single bifurcation point as drawn, some of these paths might in practice descend from higher eigenvalues of the trivial solution.

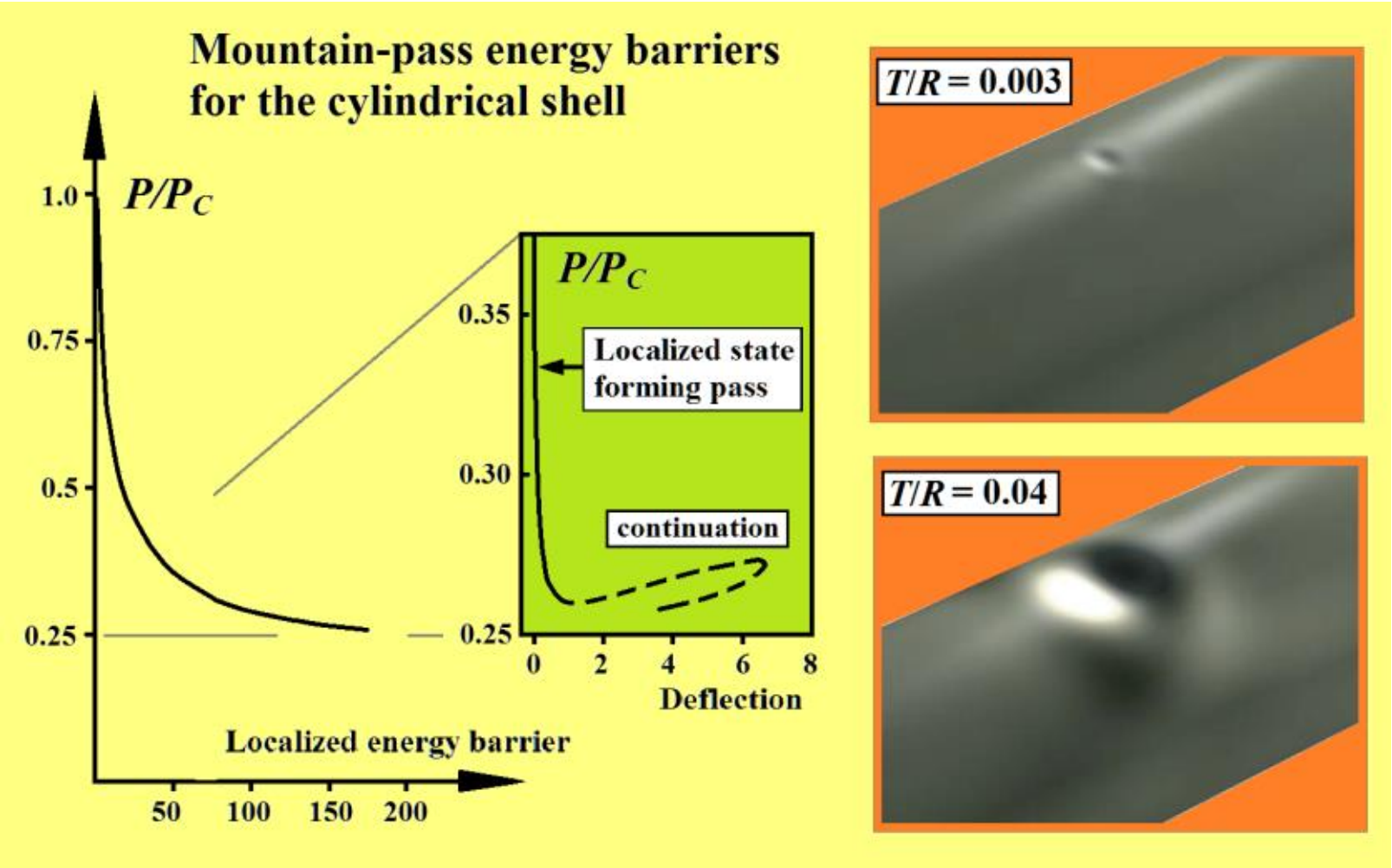

Fig 3. A summary of the results of Horak *et al* [2006] evaluating the mountain-pass energy barriers against premature escape for the axially loaded cylindrical shell.

The task of finding (for example) the lowest escape path would seem to be exceedingly difficult, but a recent theoretical paper by Horak *et al* [2006] does just that, using a mathematical search technique to locate the lowest mountain pass for a long cylindrical shell under sub-critical axial compression. The results are illustrated in figure 3 where the compressive load, $P$, is non-dimensionalized by the classical buckling load, $P_C$. The thickness of the shell is $T$, and its radius, $R$. The unstable equilibrium state of the pass has the localized form shown on the right hand side of the figure, the circumferential 'extent' of the dimple being just dependent on the thickness to radius ratio of the shell. As the prescribed value of the loading parameter, $P/P_C$, is varied, this state traces out a localized path, shown in the green thumb-nail as a load-deflection curve. Meanwhile the main graph shows the required variation of the lowest energy barrier, $E$, with $P/P_C$. This falls sharply from the bifurcation point at $P/P_C = 1$, $E = 0$, identifying the trivial state of the perfect shell as being very shock-sensitive. An objection that is sometimes raised about Horak's analysis is that it effectively allows the end supports of the compressed cylindrical shell to rotate, one relative to the other, about an axis perpendicular to the cylindrical centre line. This is indeed what allows the deflection to localize both axially *and* circumferentially. Such a rotation is rare in experimental testing, but must certainly be expected in practice where as a component of an aerospace vehicle a cylindrical shell will never be under pure axial compression, and it will be expected to carry bending moments about the perpendicular axis. This is recognized by NASA, who recently tested very large cylindrical shells under combined bending and compression [Thompson, 2015]. So, far from being at fault, Horak and his co-workers are actually being more realistic than many historical experimentalists.

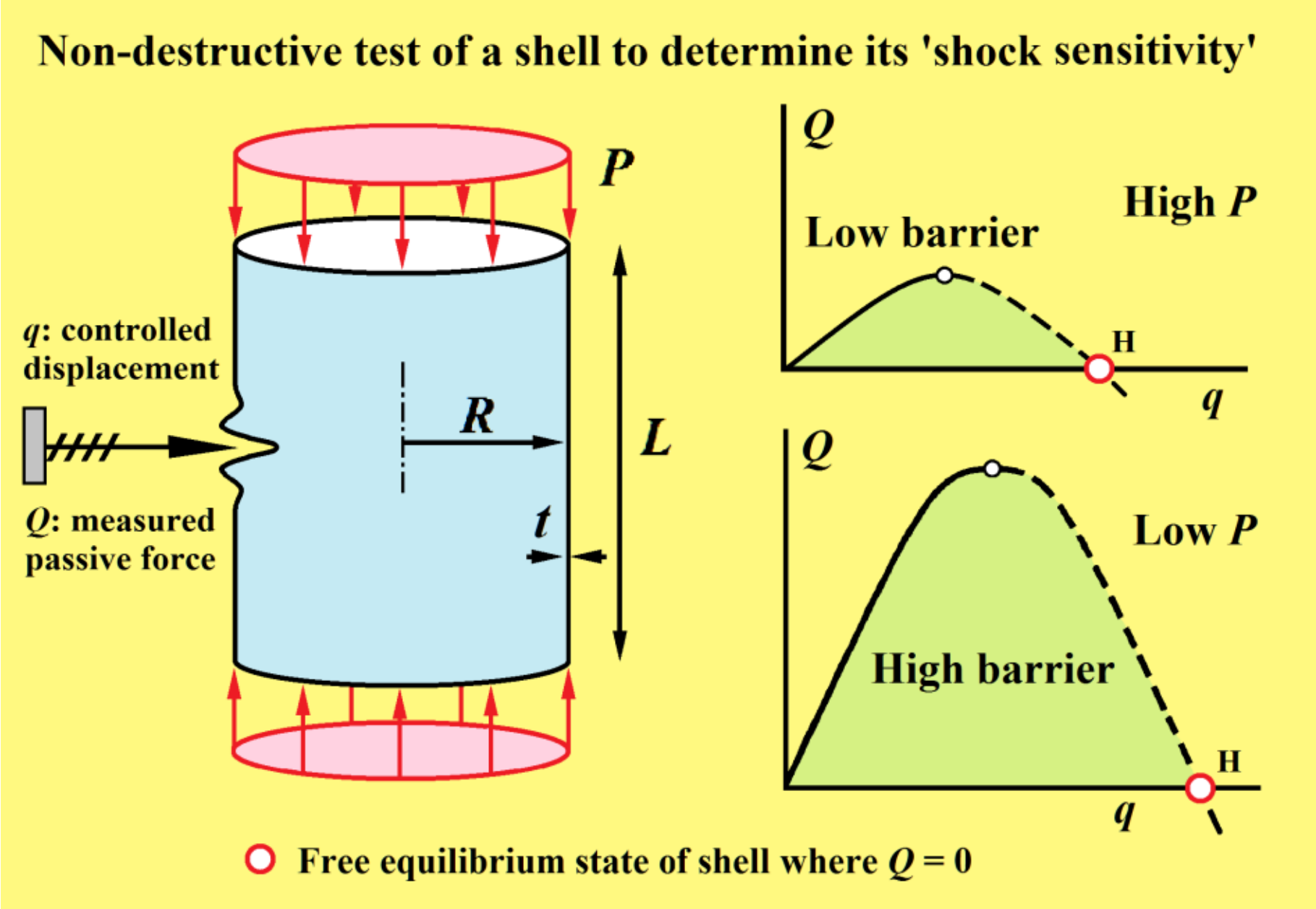

Fig 4. A schematic illustration of the proposed experimental technique as applied to a cylindrical shell. The lateral $Q(q)$ diagrams are determined at fixed values of the axial load, $P$, less than its critical value. Notional responses under two values of $P$ are shown on the right-hand side.

The localized dimple identified by Horak *et al* looks remarkably like the small dent that one might form with one's finger by pressing laterally on a small shell. This immediately suggests a non-destructive experimental procedure for locating the unstable dimpled solution (and hence its energy barrier) as sketched in figure 4. At a prescribed value of the axial compression, $P$, the idea is to use a probe with controlled-displacement, $q$, to push against the cylinder wall while at the same time sensing the passive resisting force, $Q$, of the shell. In contrast to "probing with a finger" the controlled-displacement probe will also be able to "pull", enabling it to apply a potentially sign-changing force $Q(q)$ for every $q$. With a relatively simple experimental rig, this procedure can generate in real time a lateral load-deflection graph, $Q(q)$, as shown on the right-hand side for two different values of the prescribed axial force, $P$. In the absence of any of the complications that we discuss later, each $Q(q)$ graph will have the parabola-like form shown, passing through a maximum of $Q$ and then dropping to $Q = 0$ at H where $q = q_H$. This state, H, is the free unstable equilibrium state of the un-probed shell that offers the energy barrier equal to the green area under the $Q(q)$ graph from $q = 0$ to $q = q_H$. This has been confirmed by a simulation of the cylindrical shell, and by the analysis of the spherical shell that is given in the present paper. Note that under dead lateral loading, in which $Q$ is controlled, there would be a dynamic jump from the point of maximum $Q$: but under the described rigid loading, in which $q$ is controlled, we can follow the curve to H (and beyond, if the tip of the probe is glued or welded to the shell so that it can supply either positive or negative $Q$).

## 3. How a Single Load can locate a Saddle

It might be thought that a point force on a system has very little chance of driving the system precisely to a distant saddle, but saddle points with one unstable dimension do have the property of ‘attracting’ loading paths to themselves, as we illustrate in figure 5 (in two dimensions).

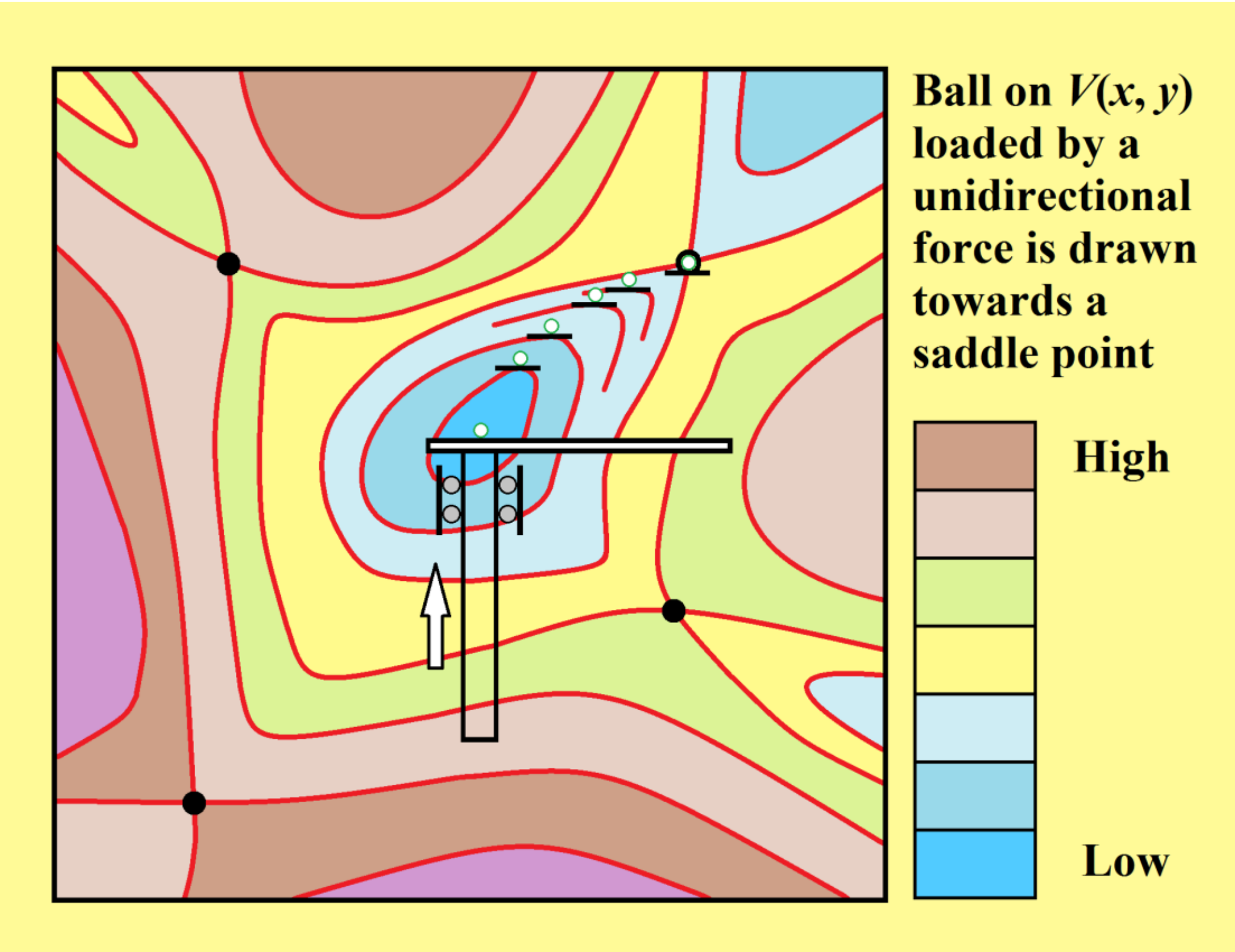

Fig 5. A ball rolling on a total potential energy landscape $V(x, y)$, viewed in plan, under a single horizontal force of fixed direction.

The figure shows randomly drawn contours of a two-dimensional energy surface with a minimum at the centre surrounded by four saddles. To think of a ball rolling on this surface subjected to a unidirectional horizontal force we can imagine a plunger pushing a frictionless plate forward, as shown. The force on the ball will always be normal to the plate, but the ball, constrained to the plate, is free to move laterally along it. This implies that the ball will follow a path on which the contour lines are touching the plate, and this has the effect of driving the ball onto one of the saddles as shown.

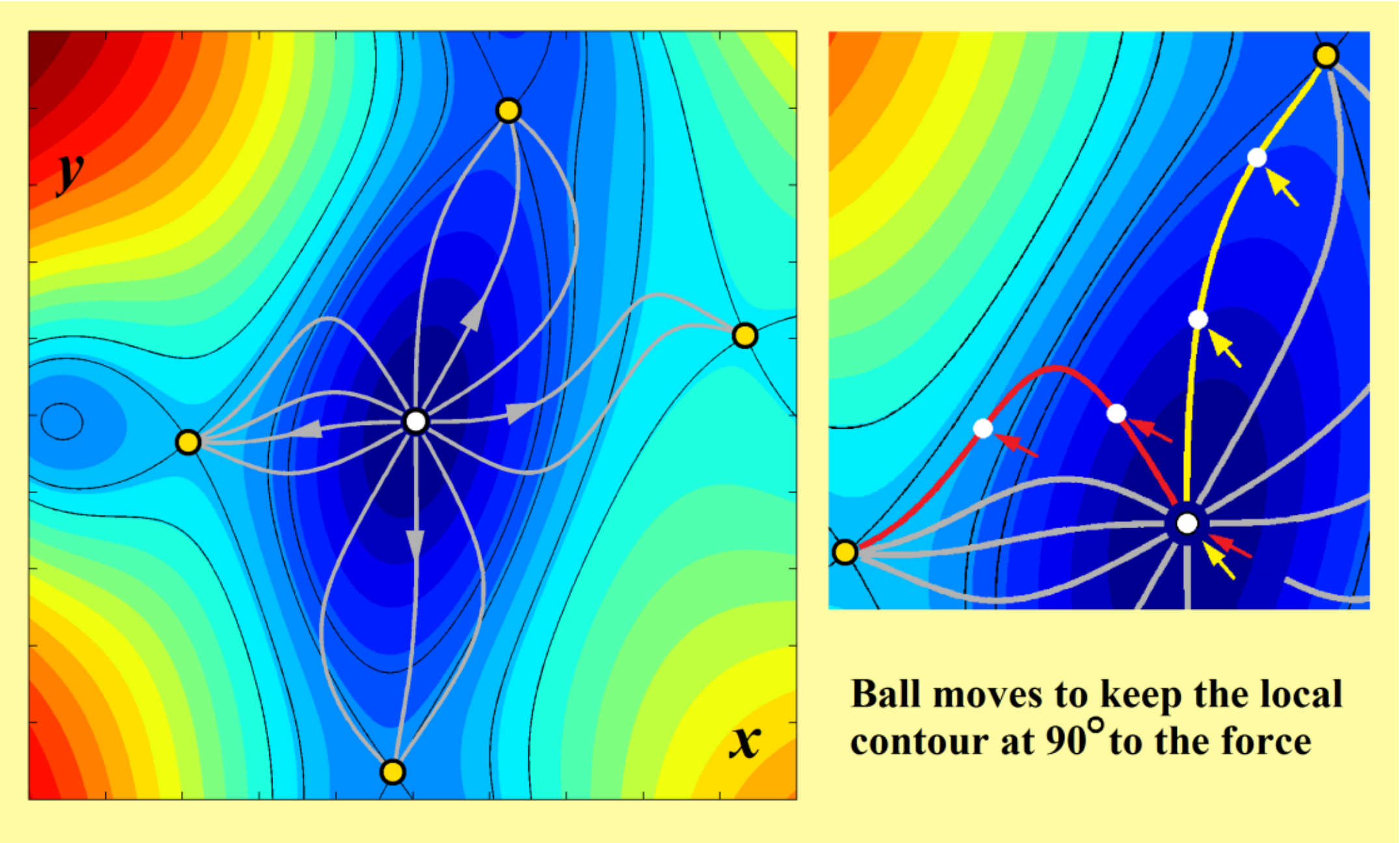

Fig 6. Computed trajectories on an analytical potential energy landscape with four saddle points, showing how all typical paths are ‘attracted’ to one or other of the remote saddles

A more complete illustration of this phenomenon is provided in figure 6, where all the paths from the central minimum have been plotted for a simple analytical energy function. The control is assumed to restrict the system rigidly to a codimension-one surface, namely to a line in two dimensions given by

$$\cos \alpha x + \sin \alpha y = c$$

where a range of fixed $\alpha$ has been chosen for figure 6 and the modulus of c is gradually increased. In this way, the system finds automatically saddles with unstable dimension 1. Thus as we vary α through 360° we find that all paths are drawn to one or other of the four distant saddle points: the only exceptions are those four atypical paths that will clearly lead to a bifurcation.

We drew attention at the end of the introduction to the idea that even a single probe will tend to pick up the lowest escape path. To elaborate on this, we consider now a situation that could arise in the buckling of a spherical shell. Analyses of Koiter and Hutchinson, that we review in section 4.1, show that the bifurcation of a (perfect) complete spherical shell under uniform external pressure involves a nonlinear interaction between an axisymmetric and an asymmetric mode, for which the simplest general model is the semi-symmetric bifurcation governed by the total potential energy, $V(x, y, p)$,

$$V = A\,x^3 + B\,x y^2 - p\,(C\,x^2 + D\,y^2)$$

Here we have two (active) generalised coordinates, the amplitude, $x$, of a ‘symmetric’ mode and the amplitude, $y$, of an ‘asymmetric mode’, while the increment of the load from its critical value is written as $p$. We shall be examining this coupled buckling with the addition of imperfections in section 4.1, but here we focus on the above perfect form. The equilibrium conditions are

$$\partial V/\partial x = x\,(3Ax - 2Cp) + By^2 = 0$$

$$\partial V/\partial y = 2y\,(Bx - Dp) = 0$$

satisfied by the trivial solution, $x = y = 0$, the symmetric solution, $y = 0$, $x = (2C/3A)\,p$, and the asymmetric solution

$x = (D/B)\,p$,    and    $[2BC\text{-}3AD]\,(D/B)\,p^2 = B^2 y^2$.

With $V$ written in the form above, the coefficients $C$ and $D$ will be positive because we want to focus on a trivial solution which is losing its stability simultaneously with respect to the principal coordinates, $x$ and $y$, as $p$ is increasing from negative to positive values. The slopes of the solutions projected onto the ($p$, $x$) plane are

Symmetric solution:    $dp/dx = 3A/2C$

Asymmetric solution    $dp/dx = B/D$

We are interested in the case in which these are both positive, implying that $A$ and $B$ are both positive, and in which the asymmetric solution has the steeper slope, so that

$2BC > 3AD$

which is also the condition allowing a real asymmetric solution in the form $y = \pm\, k\, p$ where $k = \sqrt{[\,(2BC\text{-}3AD)\,(D/B^3)]}$. Stability is determined by the second variation of $V$, and we have ensured that the trivial path is a stable minimum for negative $p$ and an unstable maximum for positive $p$. In contrast, under the prescribed conditions, the symmetric path is an unstable maximum for negative p and a stable minimum for positive p (this is akin to Poincare's exchange of stabilities in the transcritical bifurcation). Finally, each asymmetric path is everywhere an unstable saddle solution. Figure 7, with a few minor schematic changes, is based on this solution with $A = 0.5$, $B = 1$, $C = -1$, $D = -\sqrt{0.7}$ and the load level is set at $p = -1$. We shall examine this bifurcation in the presence of initial imperfections in section 4.1 (see figure 14).

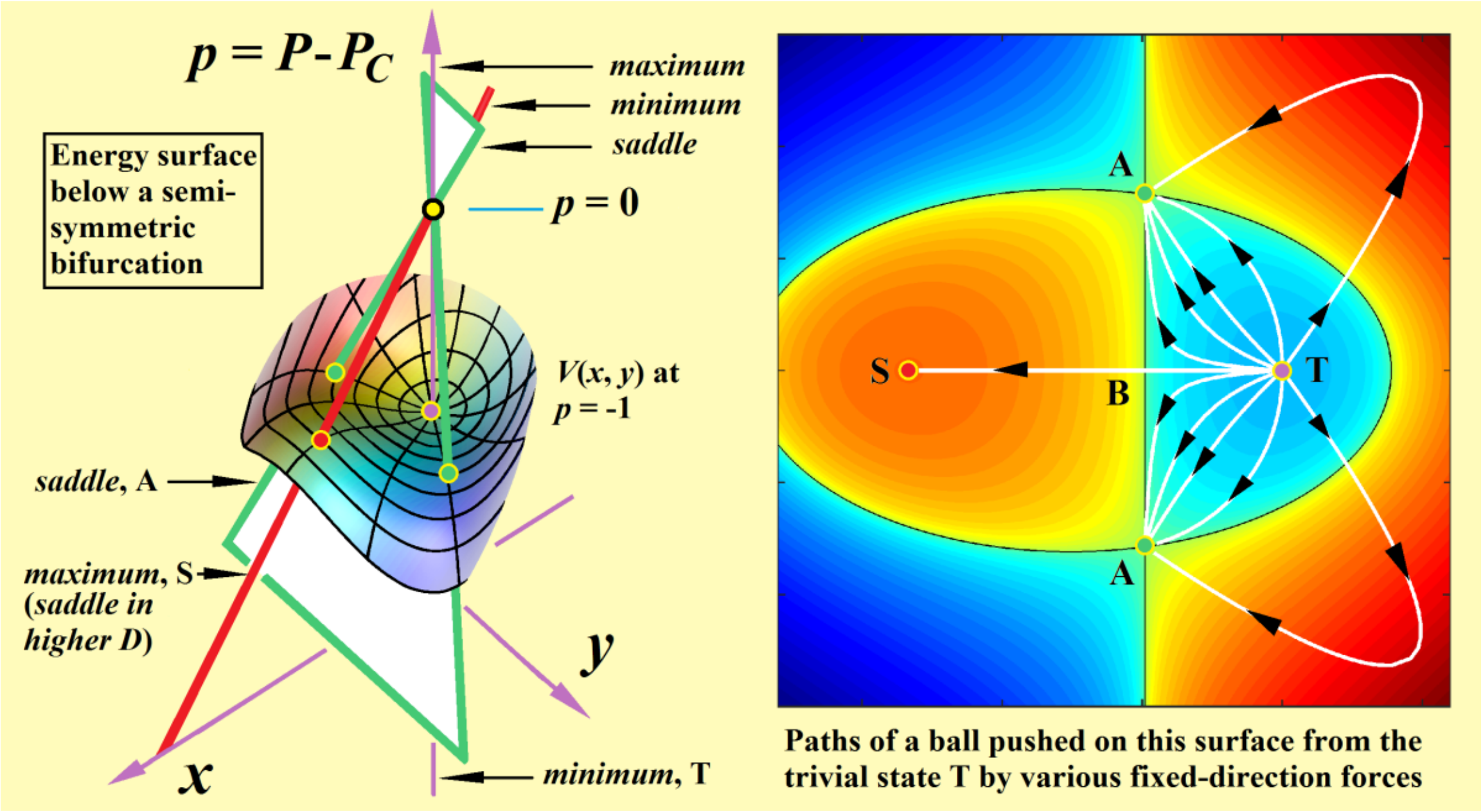

Fig 7. The normal form of the semi-symmetric bifurcation, one version of the hyperbolic-umbilic catastrophe [Thompson & Hunt, 1975]. The energy surface at a sub-critical value of $p = -1$ is displayed, and on the right-hand side the trajectories of a pushed ball are superimposed. As described in the text, the maximum, S, would be a saddle in a higher dimensional space.

On the right-hand side of the figure, the same energy surface is displayed as a contour diagram, showing the stable trivial minimum at T, the unstable symmetric maximum at S, and the two unstable asymmetric saddles at A. We should note carefully here that the maximum at S would appear as a saddle if we had drawn $V$ as a function of three or more generalized coordinates. Technically, for a multi-degree-of-freedom system, it is a saddle with two degrees of instability, while each A is a saddle with one degree of instability. Superimposed on this diagram are paths that would be followed by a ball rolling on the surface, starting at T, under a variety of fixed-direction forces, akin to those displayed in figure 6. If the force is pointed *exactly* at S, the path leads to S. This 'untypical' path clearly loses its stability at a bifurcation at B, and all other typical paths end up at one or other of the asymmetric saddles. This strongly suggests that our probing technique on a real, and therefore imperfect shell with a slightly non-symmetric energy function, will almost certainly locate the lower energy escape path over one of the asymmetric saddles.

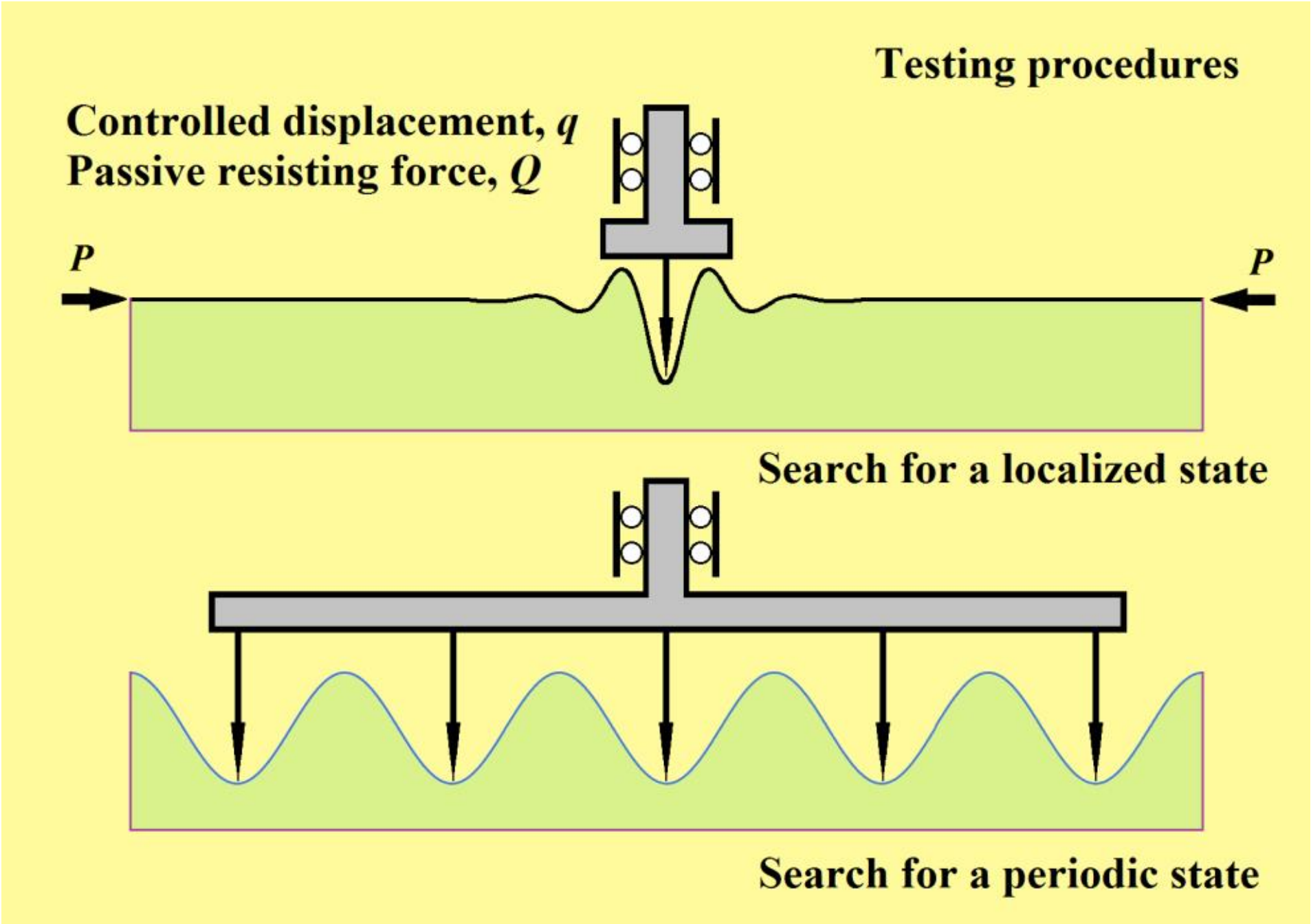

Fig 8. Two examples of how, under a single control, a search could be made for either a localized or a wavy saddle solution.

In the first part of this paper we are supposing that we are employing just a single (generalised) probe to interrogate our structure, but within this restriction there are a number of ways in which we can look for different states that we suspect might be candidates for the lowest escape saddle, H. Without going into details, one such variant is shown in the lower picture of figure 8 where a single control is applying several forces to pick up a possible periodic state. Much more variation is available if we move to hydraulic systems, but we shall not pursue this here.

## 4. Analysis of the spherical shell

### 4.1 Historical survey

***Classical linear theories.*** The initial elastic buckling instability of a complete thin spherical shell, under uniform external pressure, $P$, can be studied by a classical linear eigenvalue analysis. The first such analyses were made by Zoelly [1915] and Schwerin [1922], both of whom restricted attention to the rotationally-symmetric (axisymmetric) behaviour about an arbitrary diameter of the sphere. This restriction was lifted by van der Neut [1932] who considered general displacements (symmetric and non-symmetric) and showed that at the same critical pressure several different displacement functions are equally possible, the number of functions increasing as the thickness to radius ratio tends to zero. One of the possible solutions is always axisymmetric about some diameter, the others having a sinusoidal variation about the diameter. Thus the adjacent equilibrium form consists of a series of waves covering the whole surface, the amplitude being indeterminate. In this way it was established that the spherical shell will contract uniformly under external pressure in the normal membrane solution, which becomes unstable at the critical pressure given by

$$P_C = 2\,E\,(T/R)^2 \,/\, \sqrt{(3 - 3\nu^2)}$$

where $E$ is the Young's modulus of the material, $\nu$ is Poisson's ratio, $T$ is the shell thickness, and $R$ is the radius of the sphere.

***Early experiments.*** Early in the twentieth century, tests on the collapse of complete (or nearly compete) spherical shells are not well documented, but it was reported that experimental buckling pressures were substantially lower than $P_C$, and shells were said to jump (snap dynamically) into a localized dimple in contrast to the overall 'waving' predicted to occur at $P_C$. One set of tests by Tokugawa [1936] was published by the Japanese Society of Naval Architects. Another test, unpublished but often cited, was conducted by E. E. Sechler and W. Bollay at the California Institute of Technology in 1939. Modelling the dome of the Mt. Palomar Observatory, a brass hemisphere of 0.02 in. wall thickness and 18 in. radius was pressurized by 'immersion in a bath of mercury'. The failure pressure was close to $P_C/4$, at which the shell buckled into a small dimple. Looking ahead a little, published experimental work by Boardman [1944] and Zick & Carlson [1947] gave details of vacuum tests on large-scale storage spheres. These were essentially practical spheres, with seams, man-holes, and other discontinuities, and thus the results, while extremely useful as design data, may throw little light on the behaviour of a perfect sphere. The only vessel which was tested to the point of buckling, snapped into a dimpled configuration at approximately one tenth of the classical pressure.

***Nonlinear analyses of Karman and Tsien.*** In the same year and at the same institution as the Palomar dome test, Karman & Tsien [1939] published a seminal paper giving the first nonlinear analysis of the symmetric post-buckling states. Making sweeping assumptions to simplify the analysis, they showed that equilibrium states with large deflections exist at pressures considerably less than the classical critical pressure. Specifically, these form an unstable sub-critical post-buckling path under falling pressure, that finally restabilizes in a fold at what they termed the lower buckling load, $P_L$, namely 'the minimum load necessary to keep the shell in a buckled state with finite deformations'. They considered that while a very careful experiment may reach the critical pressure, $P_C$, all practical shells will buckle at or near $P_L$ due to initial imperfections in the practical shells, or to disturbances during testing. This work was quickly followed by Friedrichs [1941] who drew attention to the existence of a pressure, $P_M$, just a little higher than $P_L$, at which the stabilized rising path of large deflections first had a total potential energy less than that of the trivial undeflected state. This $P_M$ is the ‘energy criterion load’ emphasized (illogically) by Tsien [1942] but subsequently repudiated in [Tsien 1947]. Its true significance, in signalling the onset of ‘shock sensitivity’ (in cylindrical shells), has been uncovered in recent times [Thompson & van der Heijden, 2014]. The paper by Karman and Tsien triggered a stream of theoretical papers exploring with increasing precision the axisymmetric post-buckling behaviour of a sphere into a dimple of small central angle. Rayleigh-Ritz or Galerkin procedures were used with an assumed form of the dimple, and non-linear terms were retained in expressing the strains as functions of the displacements.

From the first nonlinear studies of the pressurized spherical shell [Karman & Tsien, 1939] and the axially compressed circular cylindrical shell [Karman & Tsien, 1941], the complex post-buckling phenomena of these two problematic archetypal buckling problems have been explored very much in tandem. Ad hoc analyses of increasing complexity were made (see for example the review by Thompson [1960a] for papers on the sphere), and it was shown for both problems that geometrical imperfections in the shape of the shell’s middle surface gave rise to a severe imperfection-sensitivity in which the bifurcation of the perfect shell was rounded off, and the failure load of an imperfect shell fell off dramatically with the magnitude of the imperfection.

***Koiter’s thesis and general theory.*** These theoretical results were put on a firm general footing by the impressive Delft thesis of Koiter [1945], written in Dutch, which in particular demonstrated the intricate nonlinear modal interactions that heightened the imperfection-sensitivity of a cylindrical shell. With the joint barriers of language and mode of publication, this work took many years to become familiar worldwide, but the thesis was eventually translated into English by NASA in 1967 (note also the recent publication of Koiter’s lecture notes edited by van der Heijden [2012]). Similar results were presented several years later by Koiter in a thorough analysis of the complete spherical shell, as we shall describe below.

***Theory and experiments at Cambridge.*** In the early sixties, the present first author completed his PhD thesis at Cambridge University on *The Elastic Instability of Spherical Shells* [Thompson, 1961], blissfully unaware (like most other researchers) of Koiter’s work as described more fully in [Thompson, 2012]. This work was focused on the complete spherical shell under uniform external pressure, and included experimental and theoretical studies. Experiments were conducted on seamless copper shells manufactured by electro-deposition on a wax mandrel [Thompson, 1960b], a technique that was followed and perfected by many researchers in the USA. The failure loads of these thin metal shells are shown in figure 9, together with the earlier results known at the time.

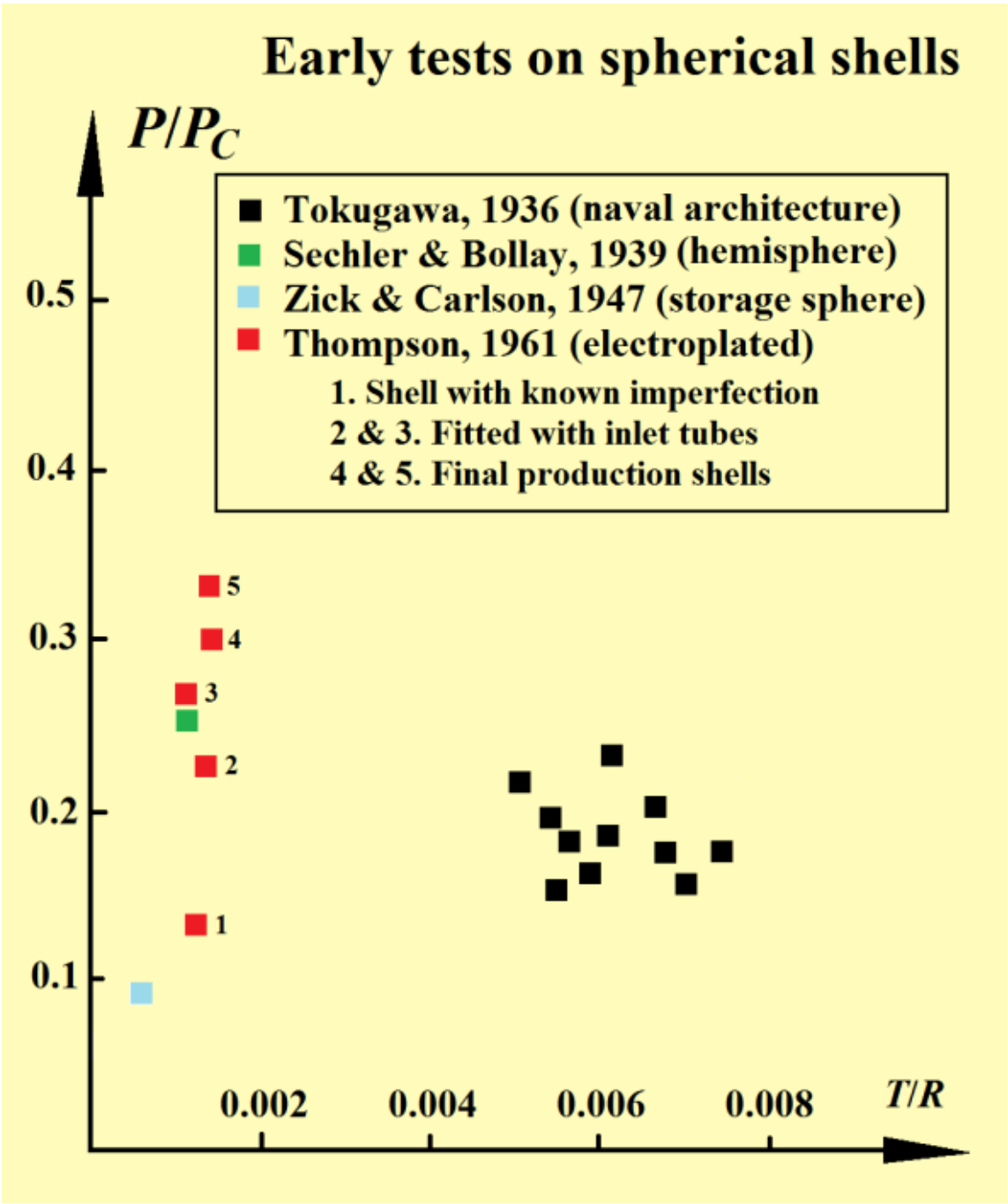

Fig 9. Experimental test results on complete spherical shells as were available in 1961. None of the experimental shells tested here had any intentional imperfections.

Photographs of two of the collapsed copper specimens (under zero pressure after much material yielding) are shown in figure 10. The first, small pentagonal dimple, formed in a dynamic jump under rigid (volume controlled) loading is similar to some elastic buckles that we shall see later. The second, large pentagonal dimple was the result of an explosive failure under dead (pressure controlled) loading. Here, as is usual, the volume control was achieved by using water both inside and outside the shell. The external water is pressurised with the equivalent of a rigidly moving piston, while the internal water remains unpressurised by means of a thin tube venting to the air. By contrast, the pressure control was achieved by using air both inside and outside the shell. The external air is connected to a large pressurized ‘reservoir’, while any internal pressure build-up can again be prevented by venting from a thin tube.

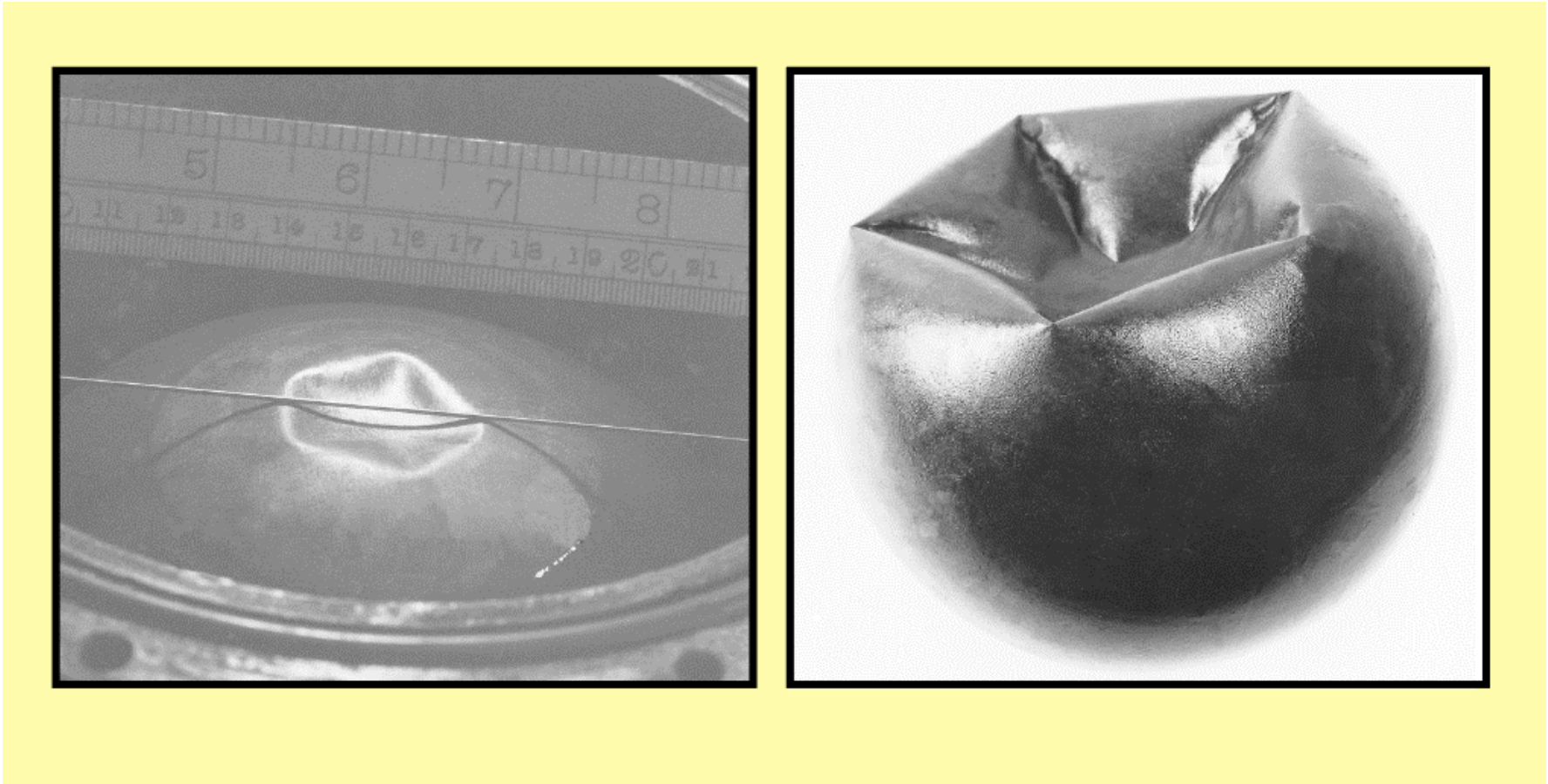

Fig 10. Pentagonal dimples in electroplated copper spheres tested at Cambridge [Thompson, 1961].

Because the dynamic jumps of metal shells, even under rigid volume-control, usually result in permanent plastic deformations, it was decided to test some relatively thick polyvinyl chloride shells to look more carefully at the elastic behaviour. These shells remained axisymmetric throughout [Thompson, 1962], and their load-deflection characteristics agreed well with a four-degree-of-freedom analysis [Thompson, 1964a] performed with the help of the Cambridge computer EDSAC 2. A follow-up paper examined the axisymmetric branching behaviour using Koiter's general theory [Thompson, 1964b].

***Sabir and Ashwell at Cardiff.*** In his work on the axisymmetric response of a uniformly loaded complete spherical shell, Sabir [1964] followed a method pioneered for point loads by his mentor Ashwell [1960]. For the purpose of analysis, a shallow region of the sphere is split into two regions as follows. An outer region effectively remains as part of the complete sphere, from which deflections are measured as usual. The inner region, meanwhile has its deflections measured from its inverted, inside-out, state. So all deflections are effectively 'small' allowing the use of linearized bending theory giving analytical results in terms of Bessel functions. Boundary conditions are satisfied to connect the two regions. The effect of a geometric imperfection was deduced from the response to a point load, and it was found that a geometrical imperfection of half the shell thickness can reduce the buckling pressure to about $P_C/5$. Experiments were made on copper and aluminium-alloy hemispherical shells, of diameter 6 in and thicknesses ranging between 0.015 and 0.020 in. Some of these had regions thinned by acid etching, giving results hopefully akin to that of a spherical shell of this reduced thickness. The best theoretical curves of pressure against the dimple-amplitude, $A$, available at this date, based on a figure of [Sabir, 1964], are shown in figure 11, including one from [Gabril'iants & Feodos'ev, 1961]. These curves are extended far beyond the reliability of the basic shell formulations, but excluding the very early curve of Karman & Tsien [1939] the important sharp fall below $A/T \simeq 2$ shows good agreement between the three latest papers. Notice that there is no clear lower buckling load, $P_L$ and consequently no clear Maxwell load (energy criterion load of Tsien), $P_M$.

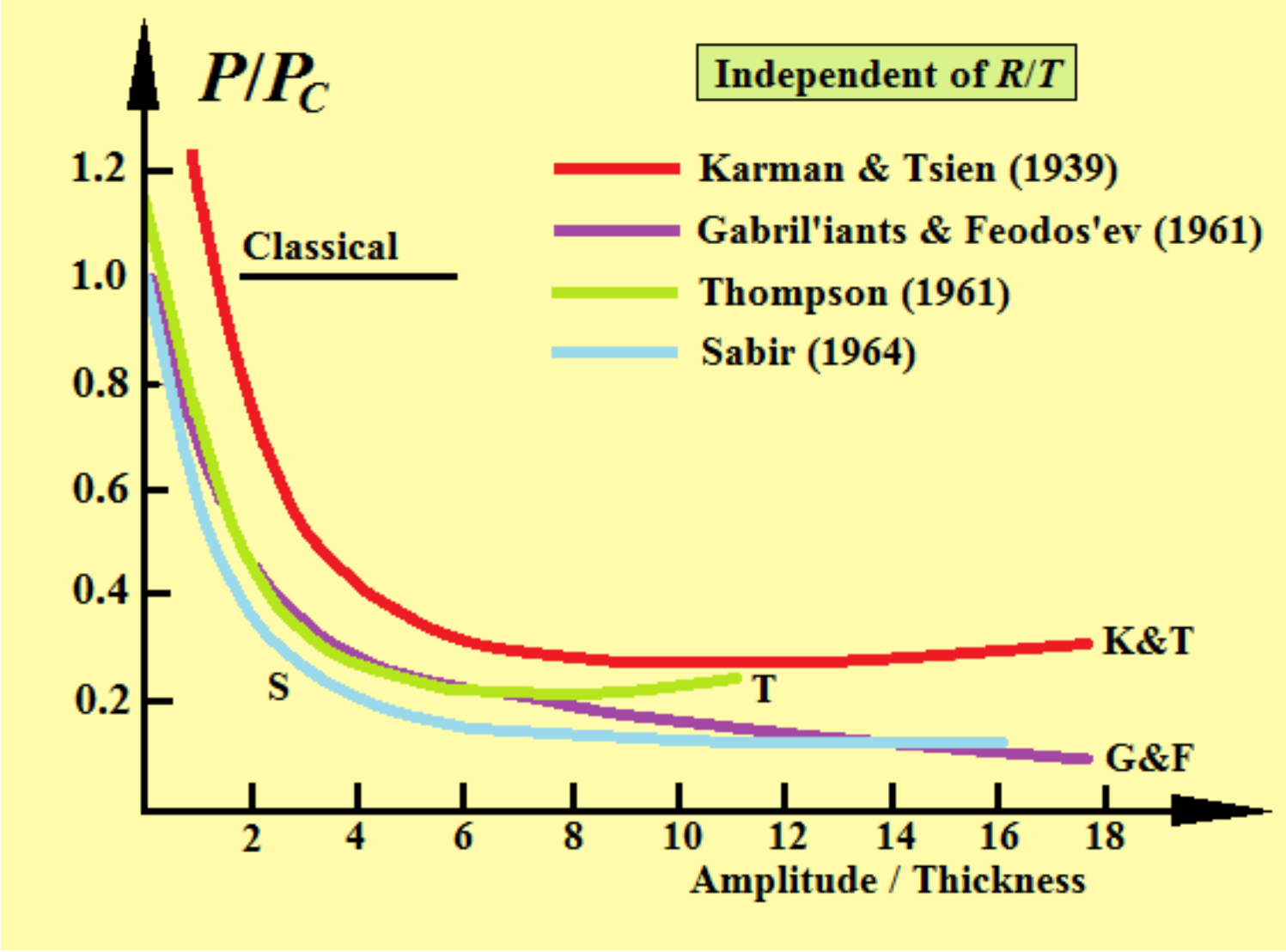

Fig 11. Non-dimensional post-buckling curves of pressure versus dimple amplitude for approximate theoretical analyses of the complete spherical shell under uniform external pressure.

***Electro-plated spheres at Stanford.*** The technique of electro-deposition was dramatically developed in the Department of Aeronautics and Astronautics of Stanford University under N.J. Hoff where the present first author spent a year as a visiting Fulbright Researcher. The paper by Carlson, *et al* [1967] describes how 32 complete spherical shells with radius-to-thickness ratios of from 1570 to 2120 were produced by electroforming. For specimens of good quality and for optimum testing conditions, buckling pressures up to 86 per cent of the classical value were obtained (though enhanced pressures were sometimes achieved by chemical polishing to produce a region of slightly reduced thickness). The effect of the loading process was examined by pressurizing the shells in rigid and dead conditions: no difference in buckling pressure was observed, effectively disproving the 'energy criterion' of Tsien [1942]. An interesting curiosity arose from dead load tests with an internal wax mandrel still inside, which allowed the formation of more and more dimples which eventually covered the whole shell surface, as shown in figure 12. This was possible because when the plated mandrel was removed from the heated plating bath the mandrel contracted more upon cooling than the nickel shell.

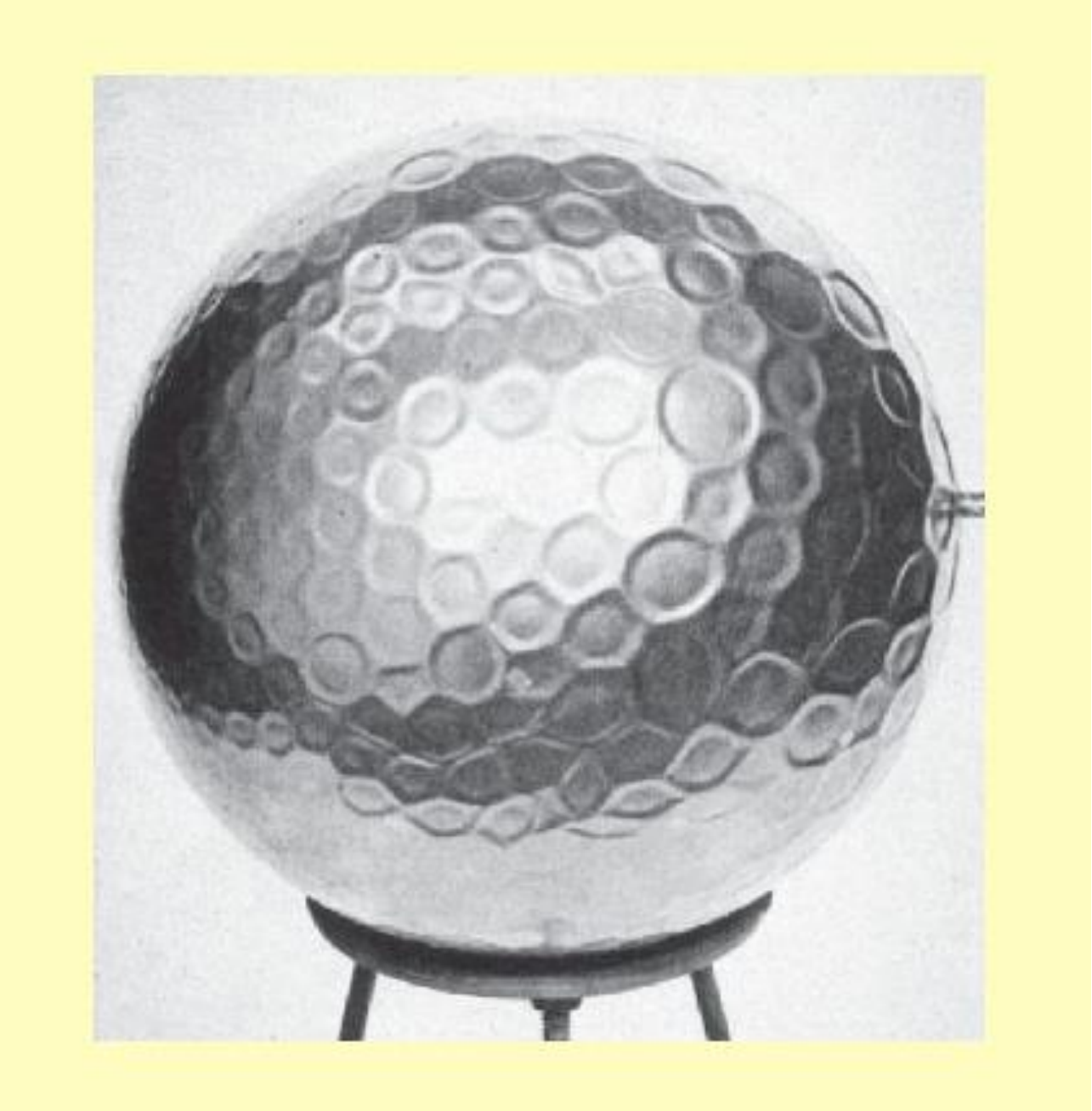

Fig 12. An electroplated nickel shell that has been continuously pressurised around a slightly smaller internal spherical mandrel [Carlson, *et al* 1967]. Some dimples are hexagonal, some pentagonal, and some circular. It could be that the non-circular dimples owe their shape to the crowding of neighbours.

This test produced the significant observation that the first dimple to hit the mandrel was sometimes approximately circular, suggesting that at initiation it might have been roughly axisymmetric.

In a follow-up paper, Berke & Carlson [1968] reported further tests on high precision electroplated nickel specimens with radius $R$ = 4.25 in. and thickness $T$ = 0.002 in., giving an $R/T$ ratio of 2125. To facilitate high speed photography they created a shallow slightly thinned area (less than 5% of the thickness was removed) to induce dimpling in a known region. Three types of test were performed:
(1) *Under rigid volume-control with the mandrel inside.* Here the mandrel does not have any effect on the initiation of the first dimple nor upon its growth until the bottom of the dimple contacts the mandrel. At this moment, further inward motion of the first dimple is restricted and strong oscillations induce transitions between several mode shapes, with often a large number of small dimples developing rapidly at other sites. High-speed motion pictures of the first inwards jump (from about 0.9 $P/P_C$) confirmed the 'common opinion' that for complete shells of large $R/T$ buckling begins with an axisymmetric inwards dimple of small central angle.
(2) *Under rigid volume-control without the mandrel inside.* With the very thin shells, and rigid loading, the authors were able to perform post-buckling tests without the mandrel which nevertheless remained in the elastic range, offering repeatable results. Figure 13 shows a typical pressure-volume result for the *unloading* of a manually induced dimple in a single shell: compare, for example, with the multiple post-buckling paths of a cylindrical shell [Hunt 2011]. The dimple starts off pentagonal in plan but repeated jumps carry it through other polygonal shapes to the axisymmetric state, and then back to the trivial path. It was noted that the transition from one mode to another was very easily activated in the neighbourhood of the crossing point of the two branches: these could be at either folds or bifurcations (into linking paths), as in similar problems [Thompson 2015].
(3) *Under dead pressure-control without the mandrel inside.* Here a high-speed movie of the initial dynamic jump shows again that the first displacement mode detected is a small *axisymmetric* dimple. As the dynamic motion continues inwards, the various static modes displayed in the figure can be fleetingly identified.

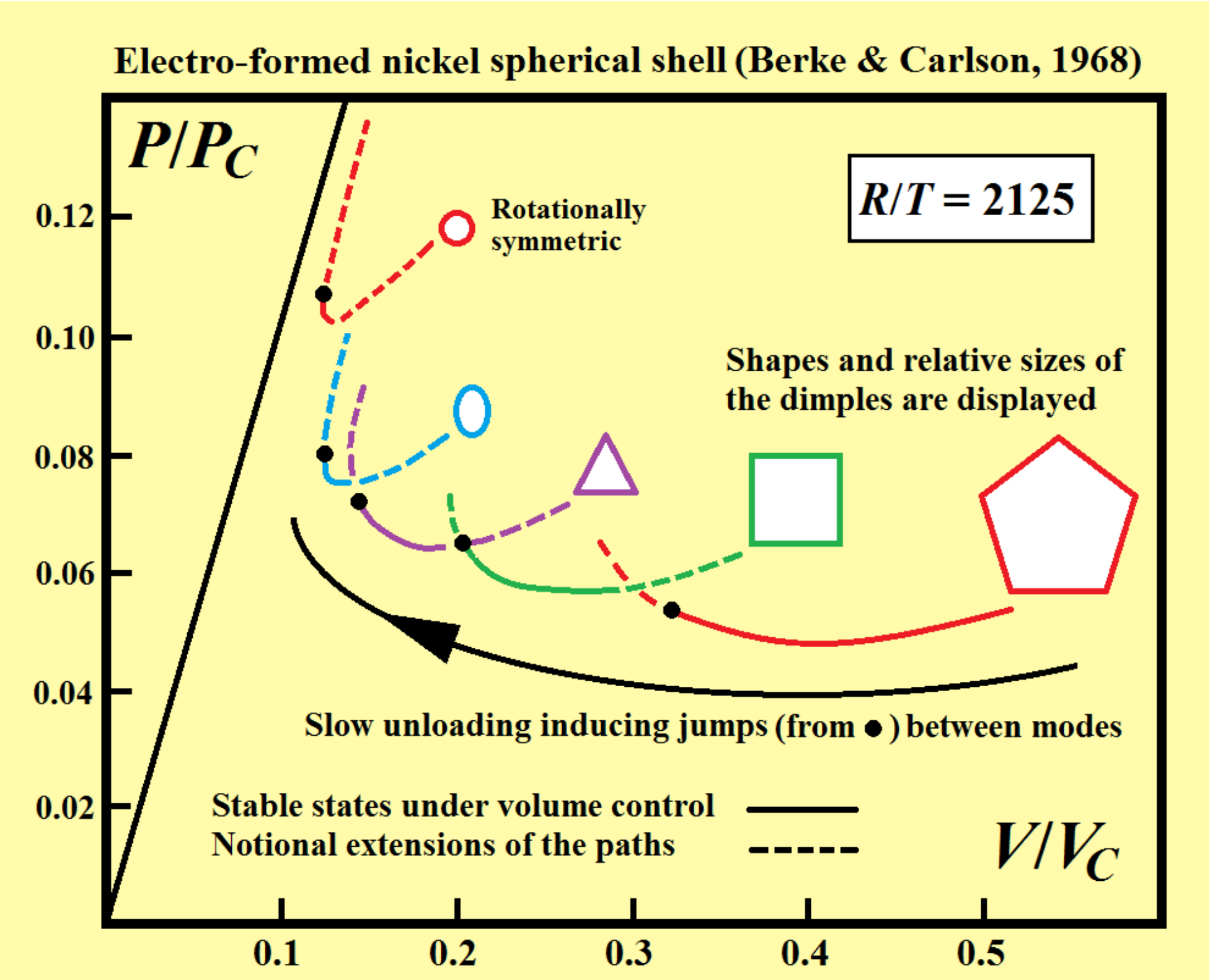

Fig 13. Results of Berke & Carlson [1968] showing the changes in dimple shape and size during an unloading sequence, on a plot of pressure versus change of volume, both scaled relative to their critical values. Note that dynamic buckling under dead load is reported as being first axisymmetric, but then transforming through the polygonal shapes, seeming to follow the reverse of the curved black arrow.

***Hutchinson's shallow coupled-mode study.*** Returning to theoretical progress, Hutchinson [1967] made an ingenious analysis of a shallow regime of a complete spherical shell via the multi-mode general theory of Koiter. He used Cartesian coordinates $x$ and $y$ chosen in the base plane of the shallow section, and the key mode into which he inserts geometrical imperfections can be written in the form $A_1 = \cos(2x)$. Although not stated explicitly, this means that we must suppose that this wave in just the $x$ coordinate corresponds to the axisymmetric mode of the complete sphere at or near to the equator, with any 'boundary' conditions ignored. He then finds that this plain rectangular wave interacts with a wave of the form $A_2 = \sin(x) \sin(\sqrt{3}y)$. With these modes, he finds a form of behaviour very similar to the nonlinear modal coupling that Koiter elucidated for the axially loaded cylindrical shell. Putting an initial geometrical imperfection in the form of the plane wave gives a curving path which intersects the plane of the coupled paths as shown in figure 14. For the sphere, Hutchinson emphasises that 'the initial post-buckling is decidedly not axisymmetric'. He also studied the coupling of 3 modes, but this seemed to give a less severe imperfection-sensitivity.

For comparison we show in this figure the similar 'homeoclinal' bifurcation discussed in [Thompson 1975, Thompson & Hunt, 1975] as one form of the 'hyperbolic-umbilic' catastrophe [Poston & Stewart, 1978]. We showed the behaviour of the perfect system in figure 7, and now in figure 14 the picture on the far right shows the fully unfolded surface of imperfection sensitivity produced by Giles Hunt for the coupled-mode buckling of a stiffened plate. Including now initial imperfections, the expansion of the total potential energy of the system about the critical point of the perfect system can be truncated to the form

$$V = A\,x^3 + B\,x\,y^2 - p\,(C\,x^2 + D\,y^2) + E\,x_0\,x + F\,y_0\,y$$

in terms of the two principal generalised coordinates, $x$ and $y$, their respective imperfections, $x_0$, $y_0$, and the increment of the load from its critical value, $p$. Here the coefficients $A$, $B$, $C$, … can of course be expressed as derivatives of $V$ at the critical point, and to generate the present three-path form of post-buckling they must satisfy the condition $2BC > 3AD$ and have $A$ and $B$ with the same sign.

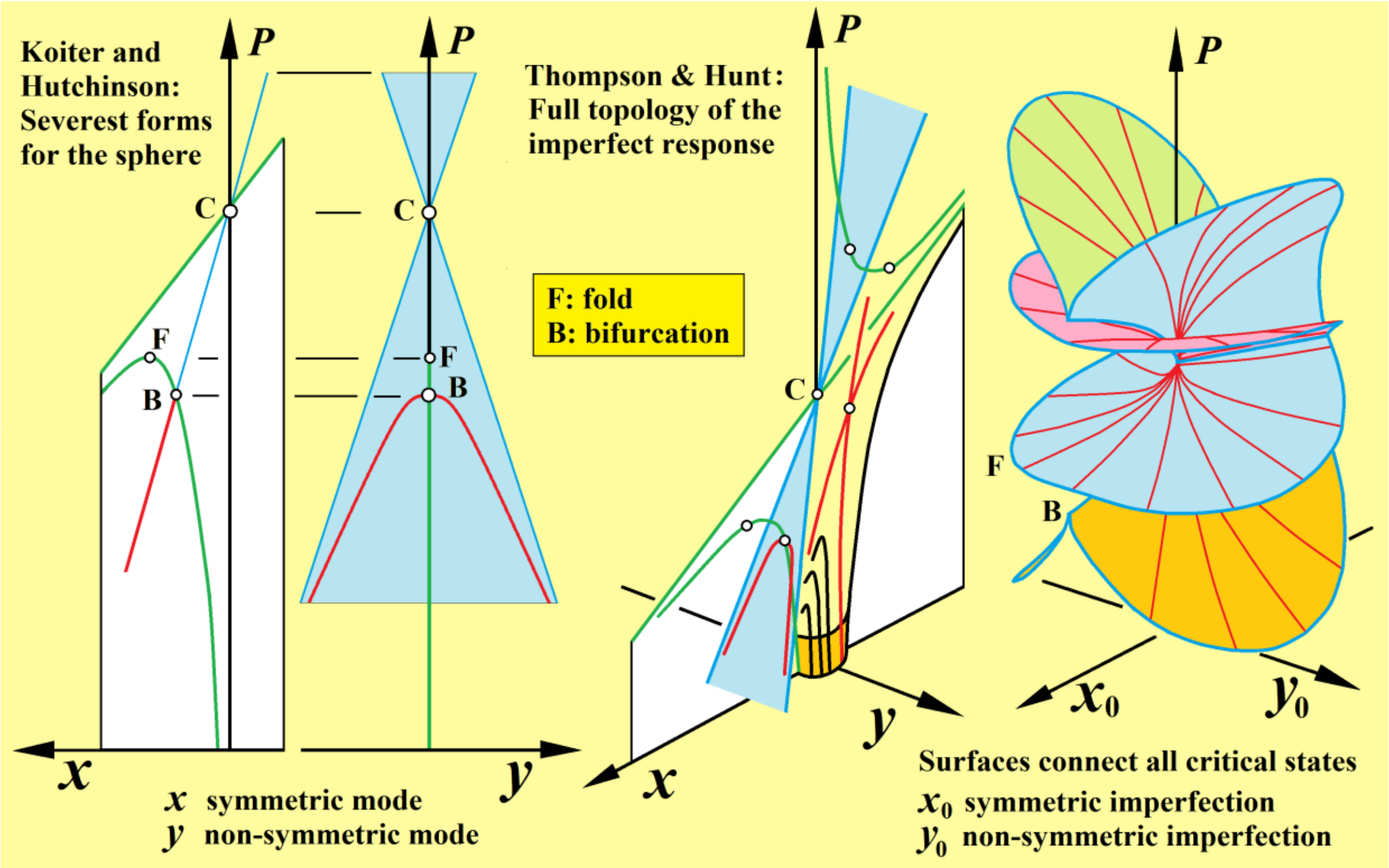

Fig 14. Nonlinear modal interactions involving a sub-critical pitchfork bifurcation from an axisymmetric imperfection path as predicted for the spherical shell by Hutchinson [1967] and Koiter [1969]. The mode of amplitude *x* has rotational symmetry, while the mode of amplitude *y* does not.

Making a comparison with the one-mode perturbation analysis of Thompson [1964b], Hutchinson found that for comparable reductions in the buckling pressure Thompson's imperfection magnitudes must be on the order of $(R/T)^{1/4}$ times the values predicted in his own study. This difference (ameliorated a little by the fourth-root!) is a manifestation of Koiter's later finding about the severe limitation of any first-order perturbation scheme. Hutchinson's coupled modes have hexagonal symmetry, as do some of the 'golf-ball' photos of the electroplated sphere repeatedly buckled onto a mandrel (figure 12). But both of these are enforced by the necessity of tiling a surface, which would be impossible with a pentagon. So it should not be concluded that a hexagon is in anyway special, or preferable to a pentagon for a single isolated dimple. Indeed figure 13 suggests that 6-sides would just lie in the sequence 3, 4, 5, 6, 7, of increasing side-numbers.

***Koiter's complete sphere analysis.*** In a massive (4 linked papers) and authoritative assault, Koiter [1969] applied his general multi-mode theory to the complete spherical shell. His first coupled-mode approximation retrieved the axisymmetric result of Thompson [1964b] and was compared in detail with Hutchinson [1967], for which some minor tweaks were suggested. It is noteworthy that Koiter's complete sphere results are dependent on $R/T$, while Hutchinson's shallow zone results are not. Finding that this first approach gave a less severe result than for the axially loaded cylinder, Koiter decided that, because of the nearby higher eigenvalues, he needed to make a more refined analysis of the axisymmetric mode, by evaluating quartic coefficients not at the critical load itself but at the actual current value of the load. He had disregarded a *systematic* perturbation extension due to its small and unknown domain of convergence. Within both his basic and refined schemes, Koiter introduced an axisymmetric imperfection, and (like Hutchinson) located a bifurcation from the symmetric path into an asymmetric mode as illustrated in figure 14. For the refined analysis he concludes that an approximately equal imperfection sensitivity occurs for purely symmetric deformations as for non-symmetric deformations, but even under the restriction to axial symmetry the post-buckling is more severe than for the cylinder. His general conclusion was that although non-symmetric modes undoubtedly play a significant role in the triggering of the buckling phenomenon, theoretical and experimental evidence suggests that in a slightly more advanced post-buckling stage the shell returns to a configuration of axial symmetry.

***Latest work.*** A paper that effectively rounded-off the stream of approximate Rayleigh-Ritz analyses of a spherical cap as a compatible component of a complete spherical shell suffering axisymmetric deformations was published by Koga & Hoff [1969]. This work used a power series expansion with a large number of free parameters for the normal deflection within the cap, and included the analysis of axisymmetric imperfections in the form of a dimple. Historically there follows a series of fairly mathematical and abstract papers focused, for example, on more accurate axisymmetric solutions. Two worth mentioning here are [Bauer *et al*, 1970] and [Graff *et al*, 1985], the latter using boundary layer theory to connect solutions between the outer shell and an inverted cap in the manner of Sabir and Ashwell.

***Symmetric or asymmetric.*** We are now in a position to summarise what we know about symmetric and non-symmetric behaviour, and we look first at the experimental evidence. First of all, we know for certain that shells with sufficiently low $R/T$ remain axisymmetric throughout, as observed by Thompson [1962] for $R/T \simeq 20$. Secondly, Carlson et al [1967] testing electroplated metal shells with $R/T$ in the range 1570 to 2120, reported that the first dimple to hit a close internal mandrel was sometimes approximately circular, suggesting that at initiation it might have been axisymmetric. Next, Berke & Carlson [1968] with $R/T \simeq 2125$ determined from high-speed motion pictures of the first inwards jump (from about 0.9 $P/P_C$) that buckling seems to begin with an axisymmetric dimple of small central angle. Further, without a mandrel, a small elastic dimple (put in by hand) adopted a pentagonal shape, but on unloading it reverted to an axisymmetric state before jumping back to the spherical state (figure 13). Turning to the theory, Hutchinson [1967] says confidently that 'the initial post-buckling is decidedly not axisymmetric'. Meanwhile, Koiter concludes that while non-symmetric modes play a role in the triggering process, it seems that in a slightly more advanced stage of post-buckling the shell returns to a configuration of axial symmetry.

So some doubts remain, but it is clear that axisymmetric studies are certainly useful, and may well capture most, if not all, of the incipient buckling behaviour. They certainly seem entirely adequate for our present theoretical study related to experimental probing for 'shock sensitivity'. However, as emphasised, the probing technique is not dependent on this.

### 4.2 Formulation of the shell analysis

As a simple analytical demonstration of how a single point load can detect a multi-dimensional saddle solution, we repeat here the four-degree-of-freedom Rayleigh-Ritz analysis of [Thompson 1961, 1964a] with the addition of a point load, $Q$ (and corresponding deflection $q$), at the North Pole of the shell. Later we shall replace the apparently 'suitable' point probe with a ring load of variable radius, to see how the procedure can cope with this when the radius is 'unsuitably' large. With only the uniform pressure load, the response is as shown in figure 15, where it is compared with tests on a relatively thick spherical ball made of polyvinyl chloride with $R/T = 19.15$ [Thompson, 1962]. Here the graph shows a plot of the applied external pressure, $P$, against the change in the external volume, $V$, both being scaled by their respective values at the bifurcation point where $P = P_C$ and $V = V_C$.

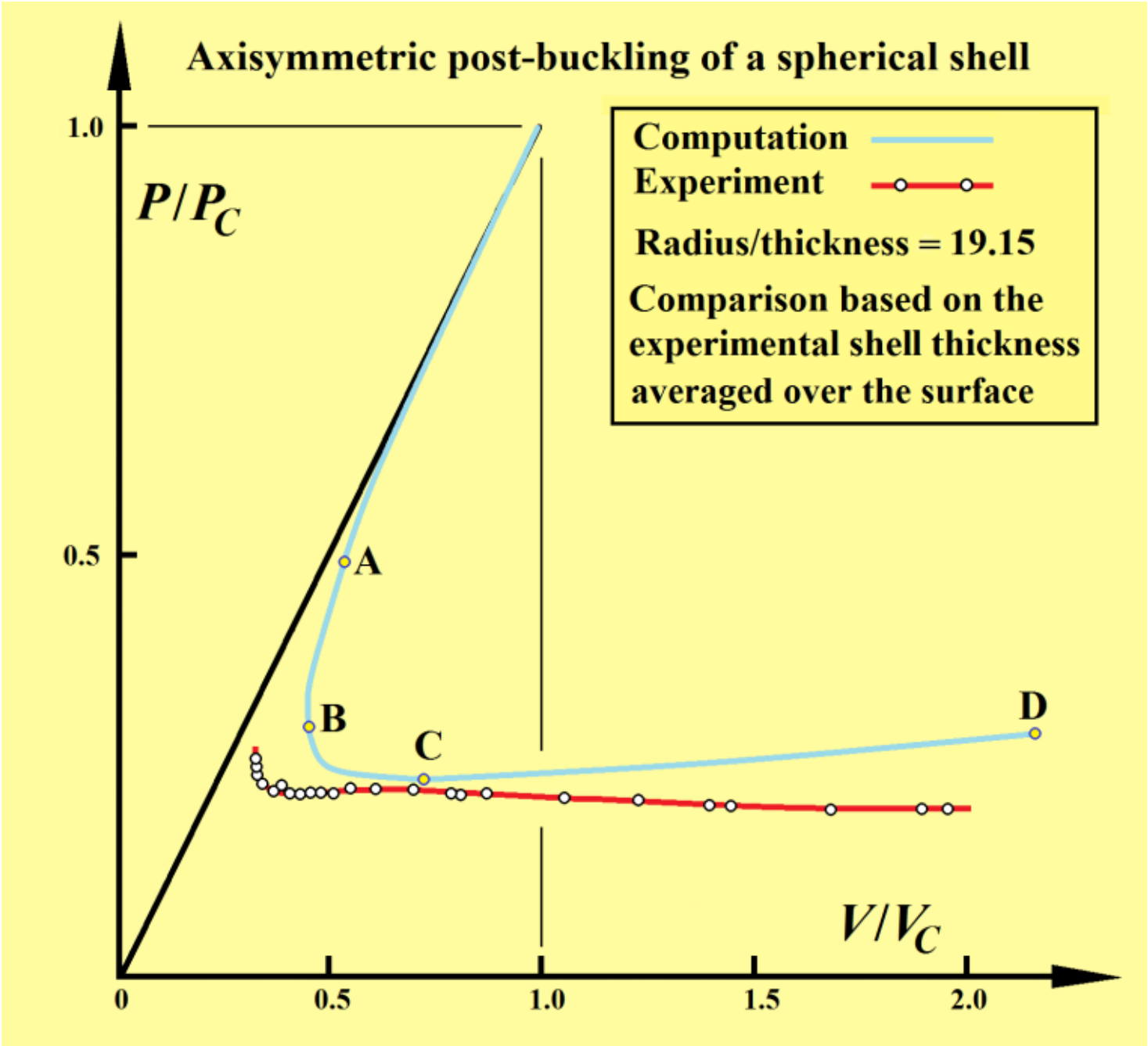

Fig 15. Axisymmetric response of spherical shells in theory and experiment [Thompson, 1962].

The notation that we use for this analysis of the elastic localized axisymmetric post-buckling of a complete thin spherical shell under uniform external pressure, $P$, is shown in figure 16.

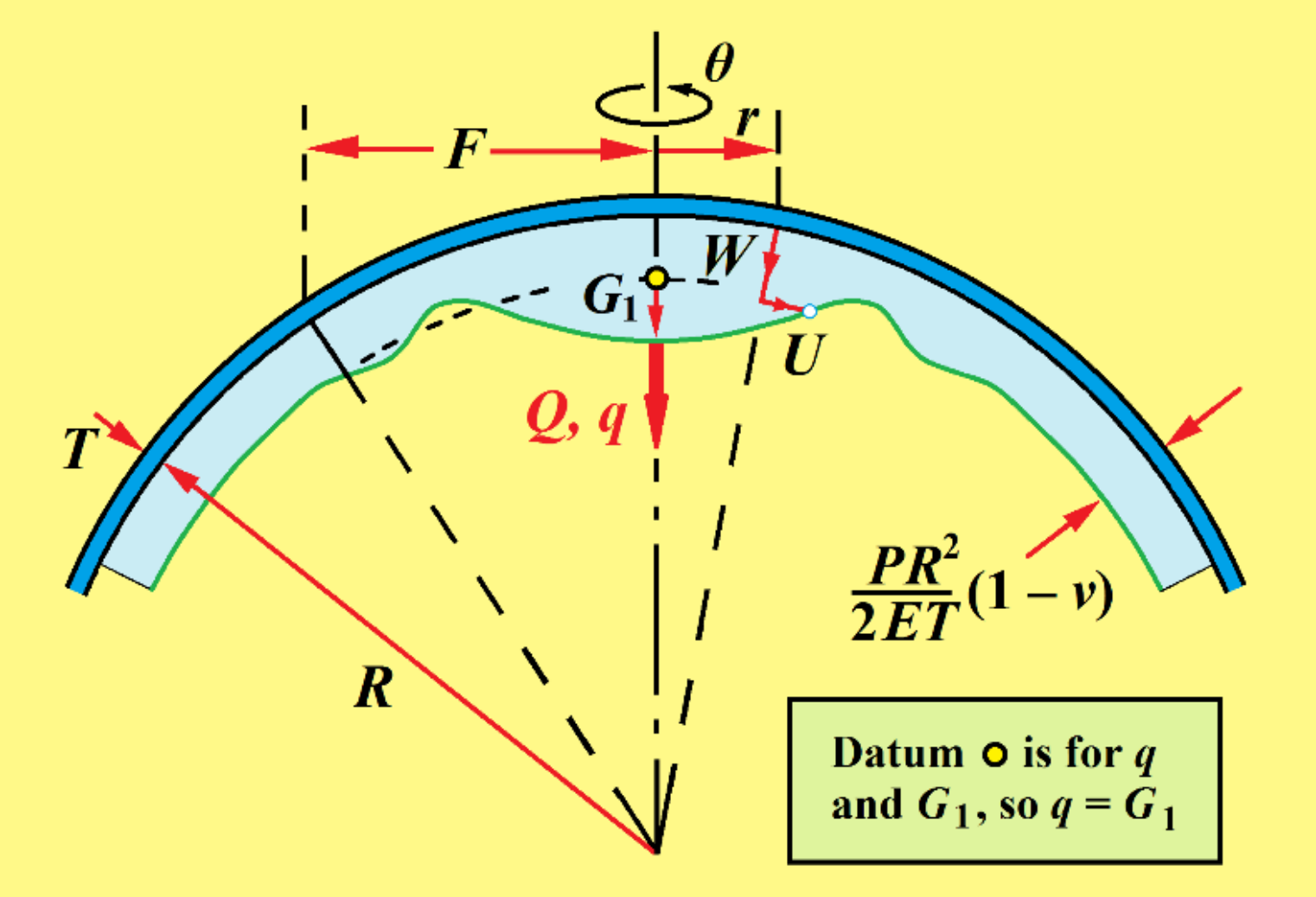

Fig 16. Notation for the analysis of a spherical shell.

The thickness of the shell is $T$ and its radius is $R$. For the shallow localized deforming regime we use as independent variables the planar polar coordinates $(r, \theta)$, so the normal and tangential displacements can be written as $W(r)$ and $U(r)$ respectively, since our assumption of rotational symmetry makes them independent of $\theta$. The localized deformation is assumed to terminate at $r = F$, where $F \ll R$. External to this localization, the shell is assumed to be in its uniformly contracted state with the constant normal deflection $W = (PR^2/2ET)(1-\nu)$ which is supplied by simple membrane theory. Here for the material of the shell the Young's modulus is written as $E$, while the Poisson's ratio is denoted by $\nu$. Compatibility with this uniform contraction at $r = F$ provides boundary conditions for the deforming cap. A four-degree-of freedom Rayleigh-Ritz analysis is made of the deforming cap for which, in common with most approximate shell analyses, the exact solution of the planar compatibility equation is used in conjunction with an assumed polynomial expansion for the normal deflection, $W(r)$. This compatibility equation for the shallow region defined by $r < F$ can be written as

$r\{(\psi r)'/r\}'/E = -(r/R)\, W' - ½\, W'^{2}$

where $W(r)$ is the normal displacement of the shell written as a function of the horizontal polar coordinate, $r$, as in the figure. The stress-function $\psi$ is the product of the radial direct stress and the radius, $r$, and a prime denotes differentiation with respect to $r$. The strain energies of bending and stretching for the shallow cap can be written as

$$J_B = ET^3/[24(1- \nu^2)]\ \ {}_0\!\int^F \{[(rW')'/r]^2 - 2(1-\nu)\, W'W''/r\}2\pi r dr$$

$$J_E = T/2E\ \ {}_0\!\int^F \{[(r\psi)'/r]^2 - 2(1+\nu)\, \psi\psi'/r\}2\pi r dr\ .$$

The assumed polynomial for the normal deflection is conveniently written, for $r < F$, as a quartic $w(\rho)$ for $\rho \coloneqq r/F$ in [0, 1] as

$$w \coloneqq W/T = (PR^2/2ET^2)(1-\nu) + g_1(1-\rho^2)^2 + g_2\, \rho^2(1-\rho^2)^2 + g_3\, \rho^2(1-4\rho^2)(1-\rho^2)^2$$

where the free parameters are $F$ and the non-dimensional mode amplitudes $g_1$, $g_2$ and $g_3$. The leading term corresponds to the uniform membrane contraction. With the chosen form for the normal deflection, we can solve the planar compatibility equation for the stress function $\psi$, and proceed to evaluate the strain energy integrals, adding in the stretching component external to the cap. Adding in, also, the total potential energy of the pressure loading, $-PV$, where $V$ is the reduction of the volume of the shell under the assumed deformation, the total potential energy, $\Phi$, of the complete shell and pressure load can be written as

$$\phi \coloneqq \Phi\, (R/\pi ET^4) = -4p^2(1-\nu)/3k(1-\nu^2) + \Sigma\Sigma\, K_{ij}\, g_i g_j + f^2\, \Sigma\Sigma\, C_{ij}\, g_i g_j$$

$$+\, p\, /\surd[\, 3(1-\nu^2)]\ \Sigma\Sigma\, D_{ij}\, g_i g_j + \Sigma\Sigma\Sigma\, F_{ijk}\, g_i g_j g_k + (1/f^2)\ \Sigma\Sigma\Sigma\Sigma\, G_{ijkl}\ g_i g_j g_k g_l$$

where $k \coloneqq T/R$, $f \coloneqq g_4 \coloneqq F/\surd(RT)$ and all summations are from 1 to 3. The new pressure parameter is $p \coloneqq P/P_C$, where $P_C$ is the classical linear buckling pressure of the real and complete *continuum shell* given by

$$P_C = 2Ek^2/\surd[3(1-\nu^2)].$$

Notice that this will not be exactly equal to the bifurcation pressure of our approximate Rayleigh-Ritz analysis for which $p = 1.15$ for $\nu = 0.33$. This may seem a poor comparison, but the localized form is chosen to be a good approximation to the advanced post-buckling behaviour, not to the initial post-buckling behaviour. The numerical coefficients, $K_{ij}$, $C_{ij}$, $D_{ij}$, $F_{ijk}$ and $G_{ijkl}$ (dependent of $\nu$) are tabulated for $\nu = 0.33$ (which we adopt in the present paper) in the Appendix. Observing that all the $g_i$, including $g_4 = f$, are valid, independently-variable generalized coordinates, we can now write the equilibrium condition

$$\partial\phi/\partial gi = 0 \qquad\qquad (i = 1 \text{ to } 4) \qquad\qquad (1)$$

and solve for the $g_i$ in terms of the pressure parameter, $p$, and Poisson's ratio, ν. Notice that the thickness/radius ratio, $k$, which appears only in the leading (constant) term of $\phi$, will not be involved, but will re-appear if we need the volume parameter which we write as

$$v \coloneqq V/V_{UC} = p + k\, \lambda\surd[3(1-\nu^2)]/(1-\nu)$$

where the useful dimple-volume parameter, $\lambda$, (independent of $k$) is given by

$$\lambda \coloneqq ½\, f^2\, [\, g_1/6 + g_2/24 - g_3/40\, ].$$

Here $V$ is the total volume change, and $V_{UC}$ is its value in the uniformly compressed membrane state at $P = P_C$. The results of the analysis so far (with no point load at the pole) simply duplicate the results of the thesis [Thompson, 1961] that are displayed in figure 15. This shows the dependence between $V/V_C$ and $P/P_C$ for stationary values of the potential energy, $\phi$, given by solutions of equation (1).

### 4.3 Addition of the point probe

To study the extra deflection of the pressure-loaded shell produced by the addition of a point load, $Q$, at the North Pole, we can treat $q \coloneqq G_1$ as the corresponding deflection (because the pressure $P$ will always be held constant during the probing). So to perform the point-probe analysis we must augment the energy expressions (dimensional and non-dimensional) by

$$\Phi^* = -Qq$$
$$\phi^* = -Q_S g_1$$

where the scaled point load parameter is defined by

$$Q_S \coloneqq QR/(\pi E T^3).$$

We can now solve the equilibrium equations (1) for the augmented potential energy, the results being again independent of the thickness/radius ratio, $k$, unless we want to assess volume changes.

### 4.4 Addition of the ring probe

As an alternative to the point probe at the pole, we now study an axisymmetric ring probe. For this it is convenient to imagine a ring of conical forces normal to the undeflected shell of total magnitude, $Q$, attached to material points of the shell at the ring radius where $r = r_R$ as shown in figure 17.

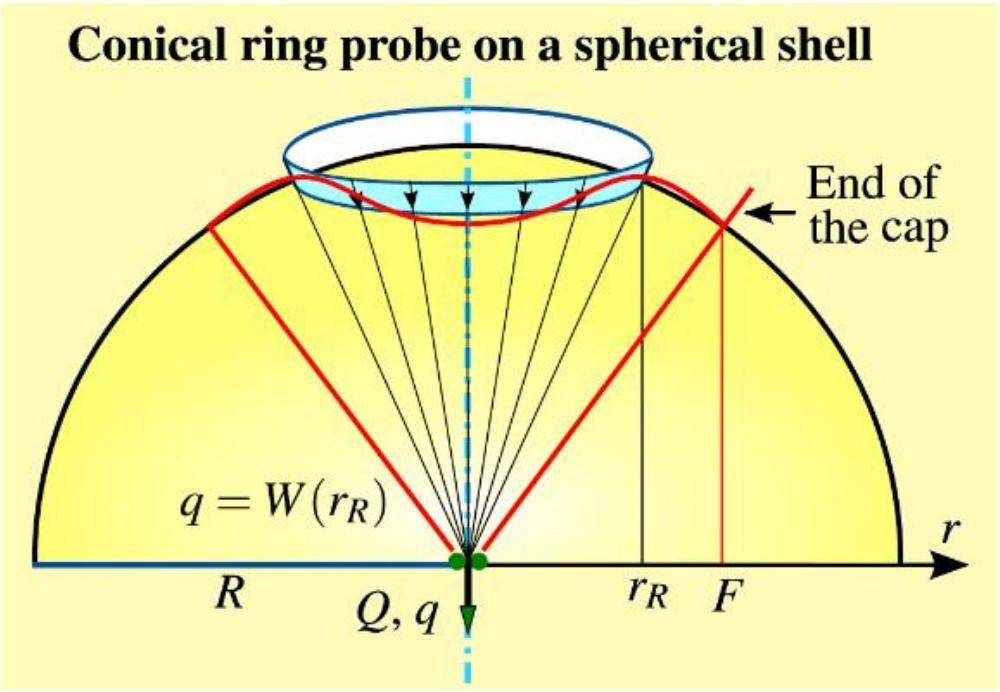

Fig 17. Notation for the sphere with a conical ring probe, illustrating a case where we would not expect the method to work well (if at all) since the shell wants to move outwards at the ring radius.

These forces, which are assumed to maintain their magnitude and direction as the shell deflects, could be provided (easily as dead loads, less obviously as rigid loads) by a system of strings passing through pulleys at the centre of the sphere as illustrated in figure 17. The corresponding displacement of the generalised load, $Q$, is $q = W(r_R)$ or in non-dimensional terms $q_S \coloneqq q/T = w(\rho_R)$. To analyse this arrangement, we need to *augment* the energy of the system, before solving the equilibrium equations (1), by

$$\Phi^{**} = -Qq = -QW(r_R)$$
$$\phi^{**} = -Q_S q_S = -Q_S\,[g_1(1-\rho_R^2)^2 + g_2\,\rho_R^2(1-\rho_R^2)^2 + g_3\,\rho_R^2(1-4\rho_R^2)(1-\rho_R^2)^2\,]$$

where $\phi^{**}$ is a displayed function of the generalized coordinates ($g_1$, $g_2$, $g_3$) and an implicit function of $g_4$ via the definition of $\rho_R \coloneqq r_R/F$. Here, as before, the scaled load is $Q_S \coloneqq QR/(\pi E T^3)$ and since $g_4\sqrt{(RT)} = F$, we have $\rho_R = r_R/(g_4\sqrt{(RT)}) = \alpha/g_4$, introducing the dimensionless ring-radius parameter $\alpha \coloneqq r_R/\sqrt{(RT)}$. The two non-dimensional parameters representing the radius of the load, $\alpha$, and the radius of the deforming cap, $g_4$, can be written respectively as

$\alpha = (r_R/R)\,\sqrt{(R/T)}$

$g_4 = (F/R)\,\sqrt{(R/T)}$

with $\alpha\,/g_4 = \rho_R$ .

Now we clearly must have $\alpha\,/g_4$ less than unity, but this is taken care of by the energy minimization process, which will always choose a value for $g_4$ greater than $\alpha$, so that a fall of the load is possible. Another concern is that the deforming cap should always remain shallow, as required by our method of analysis. But for any determined values of $\alpha$ and $g_4$, the real angular measures, $r_R/R$ and $F/R$ can always be made suitable low by decreasing $k = T/R$ (which, as before, does not enter the analysis). The deflections will always be shallow for a shell that is thin enough relative to its radius.

## 5. Computational demonstrations for the spherical shell

The results of our computations in sections 4.3 and 4.4 can now be used as simulated experimental tests of our probing technique.

### 5.1 Response under the point-probe

The results of the point-load study are summarized in figure 18. The three-dimensional diagram at the top left shows plots of $Q(q)$ at various values of the load parameter $p$. They all rise up from the back starting line, pass through a maximum of $Q$, and then descend to $Q = 0$ on the base curve in the plane of $p$ against $q$: which is, as expected, the known post-buckling path of the unstable saddles of the free shell, H, that form the energy barriers. This confirms the scenario of our shock-sensitivity determination illustrated in figure 4.

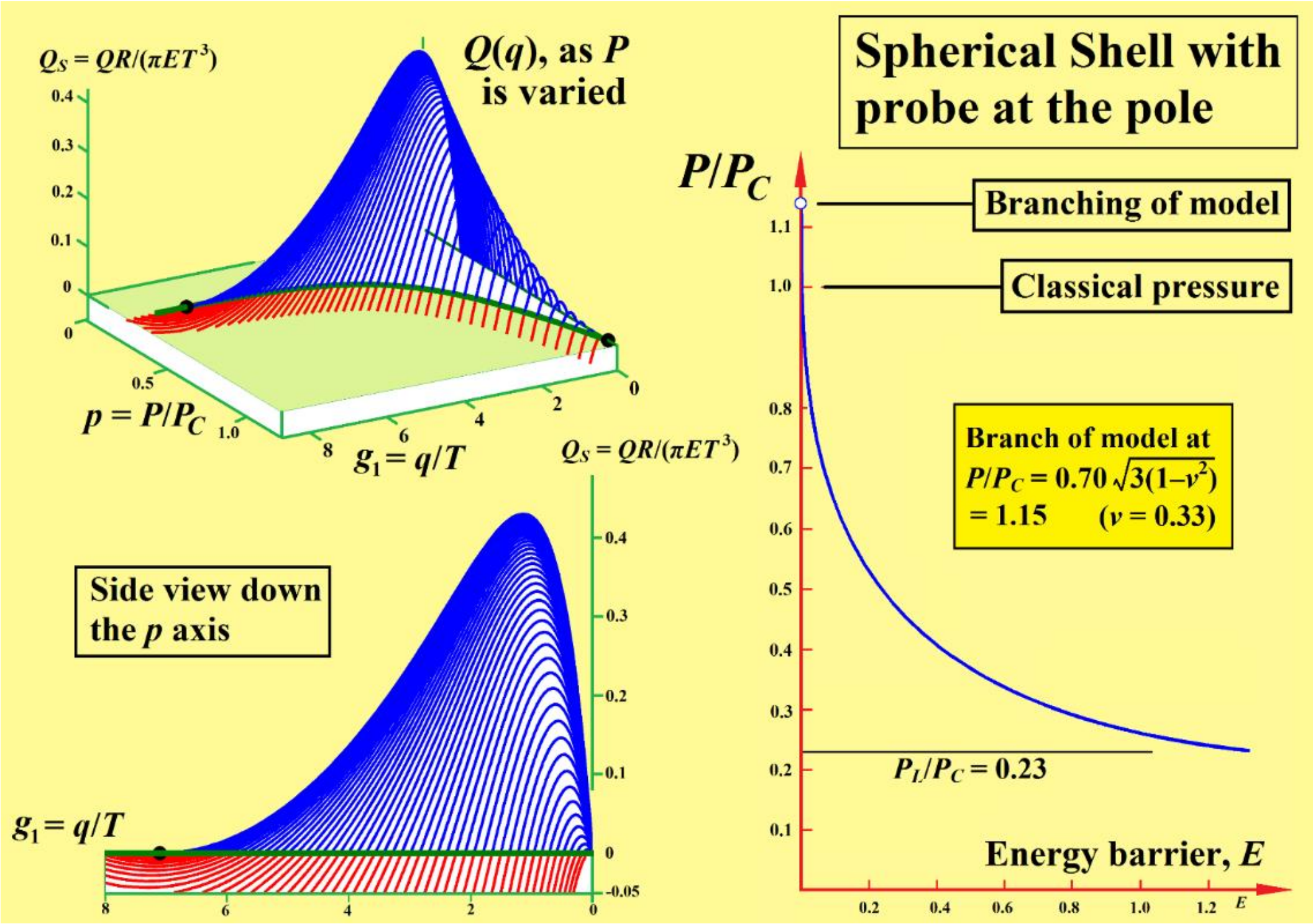

Fig 18. Analytical results with a point probe at the pole of the spherical shell. The right-hand graph of shock sensitivity on a plot of $P/P_C$ against the energy barrier, $E$, is the exact answer (within the present theoretical model) because the point probe has everywhere located the free post-buckling saddle, H.

The bottom left diagram is simply a view of the three-dimension surface looking down the $p$ axis, to show the true shapes of the $Q(q)$ curves. Meanwhile the right-hand graph shows how the energy barrier, $E$, supplied by the areas under the $Q(q)$ curves, varies with the uniform external pressure on the shell, represented by $p = P/P_C$. Notice that, as with the cylindrical shell of figure 3, the $p(E)$ curve drops sharply from the bifurcation point, revealing a similar severe form of shock sensitivity.

**5.2 Response under the ring-probe**

The results for the low value of $\alpha = 0.5$ are shown in figure 19. We see that the graphs of $Q(q)$, here in their non-dimensional form as $Q_S(q_S)$, have the simple form that we have already demonstrated for the shell under a point load. The curves increased monotonically with $q_S$ with no vertical tangencies. They therefore locate in a straightforward manner the free post-buckling states of the uniformly compressed spherical shell shown in dark green in the plane of $Q_S = 0$. The areas under the graphs then give us the energy barriers against premature buckling which agree exactly with the (correct) answers found by the point-probe analysis.

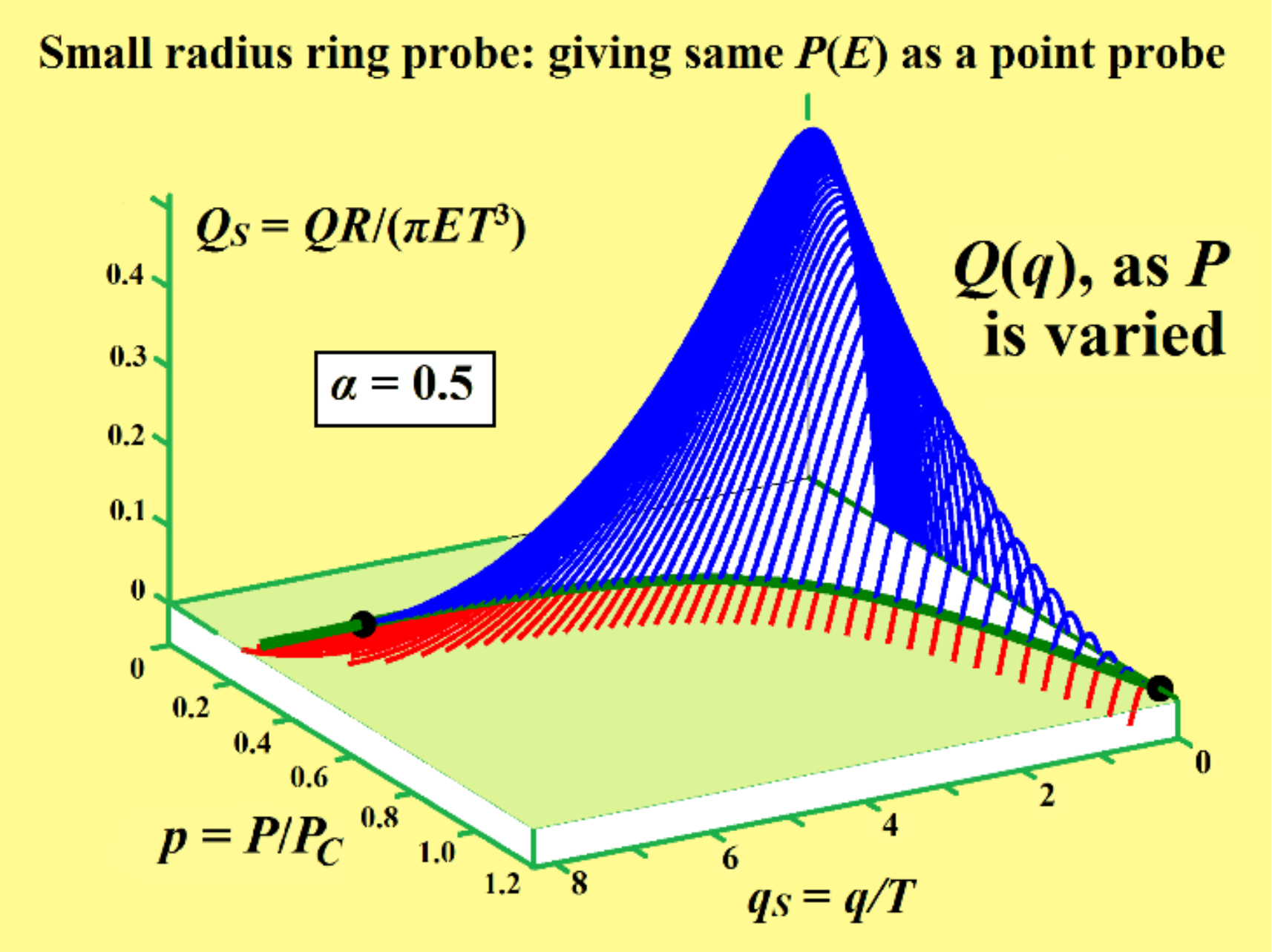

Fig 19. The sphere under a ring-probe of small radius ($\alpha = 0.5$), behaving qualitatively like the shell under a point probe. The shock-sensitivity predicted is identical to that shown in figure 17, because, once again, the scheme has accurately determined the free post-bucking solution, H.

To see how and why this behaviour changes at higher values of $\alpha$, we need now to compare the possible ring radii to a 'typical' post-buckling form, and we must start this comparison by looking at figure 20. This shows the growth of the unstable post-buckling dimple of the free shell as $P$ is decreased to $P_L$, and in particular its form at a value of $P/P_C = 0.33$ which we have chosen for further study.

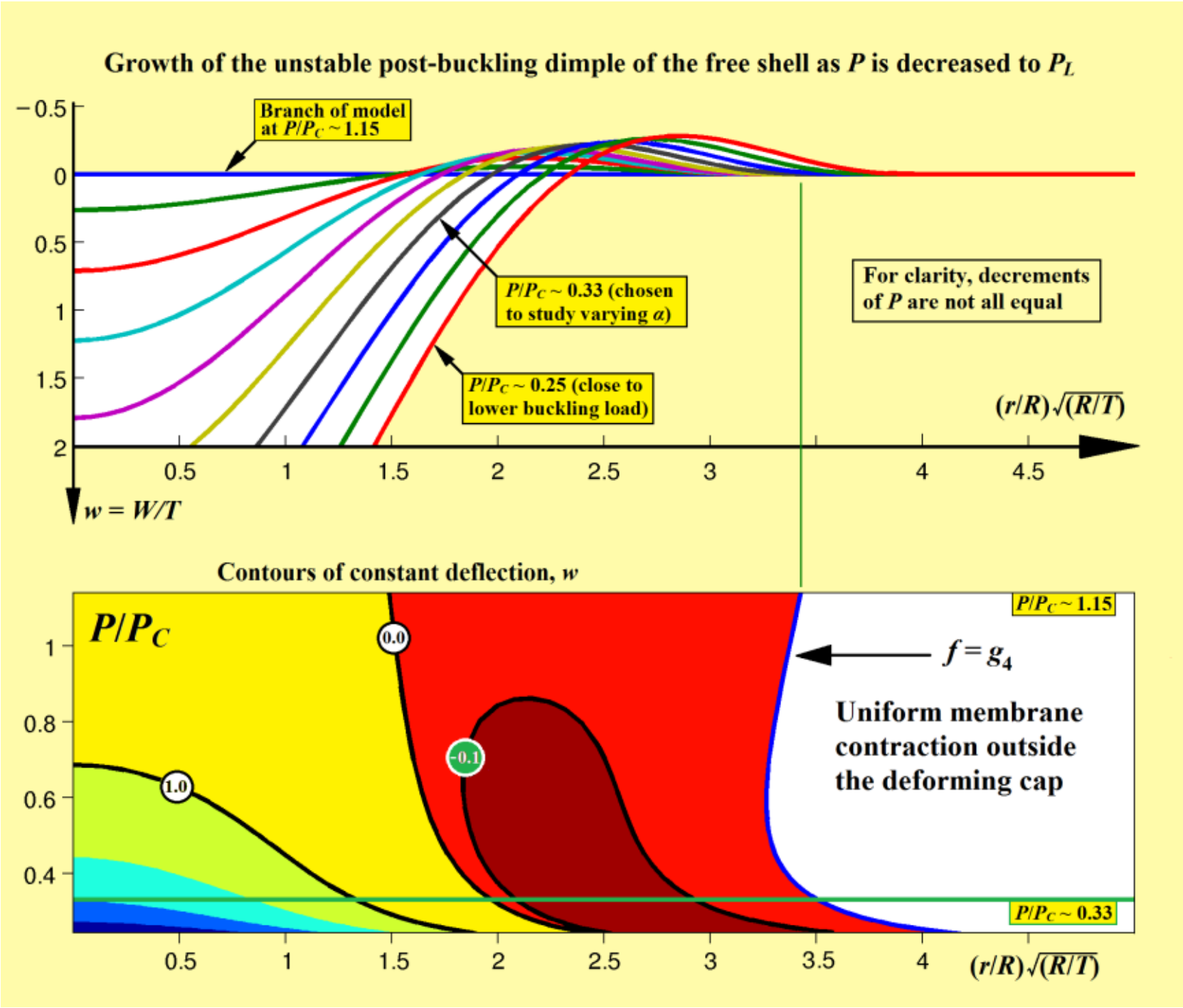

Fig 20. The unstable post-buckling forms (H) of the free un-probed spherical shell. The top diagram shows the dimple shapes at various $p$ levels on a plot of the non-dimensional normal deflection, $w = W/T$ against the radius parameter $r/\sqrt{(RT)}$ . The lower diagram shows the same information as contours of $w$ over the plane of $p$ versus $r/\sqrt{(RT)}$ This figure was drawn to help establish appropriate values for $p = P/P_C$ and $\alpha$ for our following studies

For this value of $p = P/P_C$ we reproduce the post-buckling form in figure 21, together with the ring positions that we shall now investigate at $\alpha = 0.5$ to 2.75 in increments of 0.25. Those shown in blue are clearly ‘inappropriate’ since the ring forces are pressing down where the sphere wants to move upwards. It is this behaviour, and how it develops through the intermediate positions shown in red, that we wish to understand.

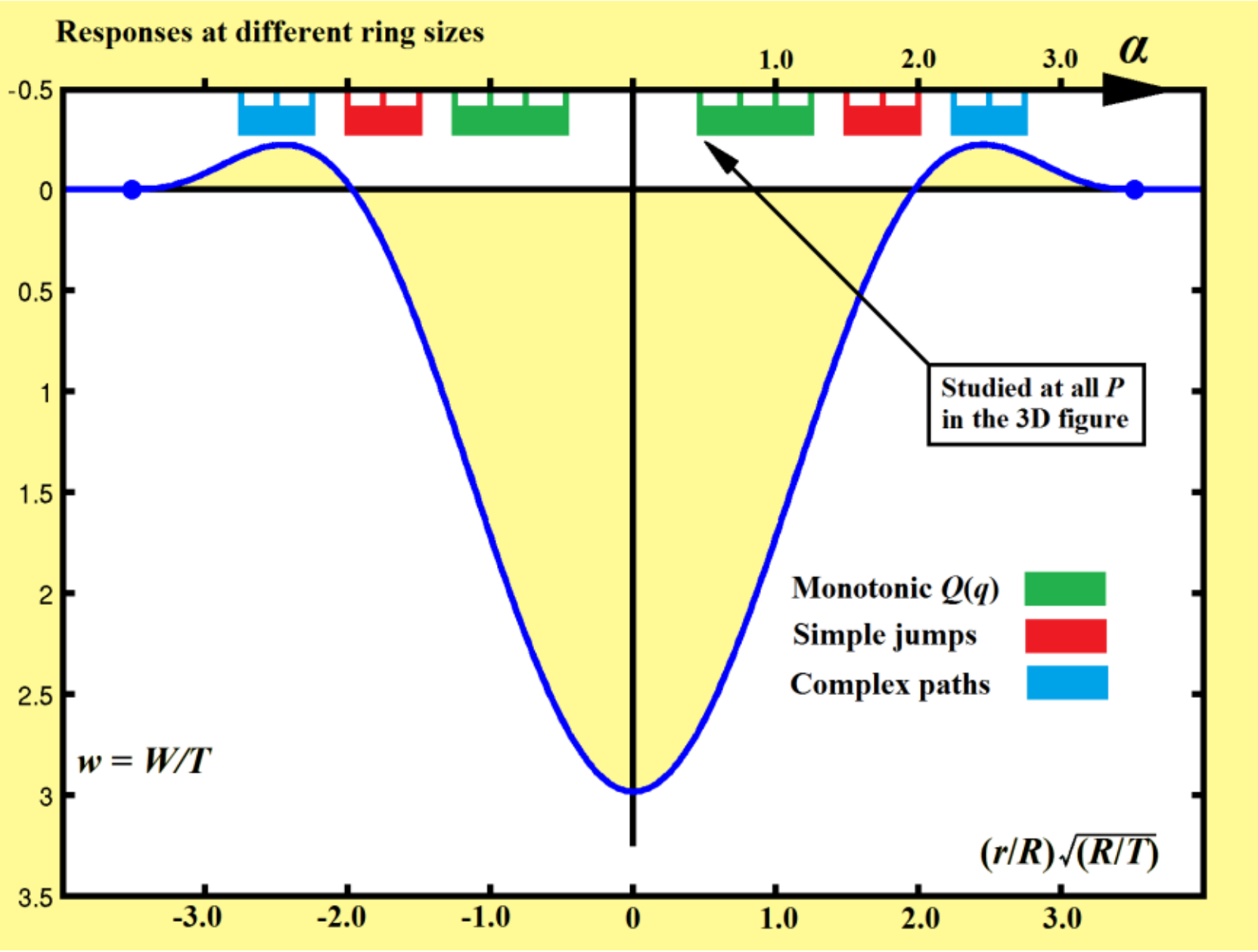

Fig 21. Positions of the ring probes defined by the coloured values of $\alpha$. These were selected as challenging positions in relation to the post-buckling saddle, H, at our chosen load level, $p = P/P_C = 0.33$.

Figure 22 shows the low $\alpha$ range, where the behaviour is akin to that of the sphere under a point load. The free states of the un-probed shell, H, are accurately located, allowing the energy barrier to be evaluated as the area under the curves.

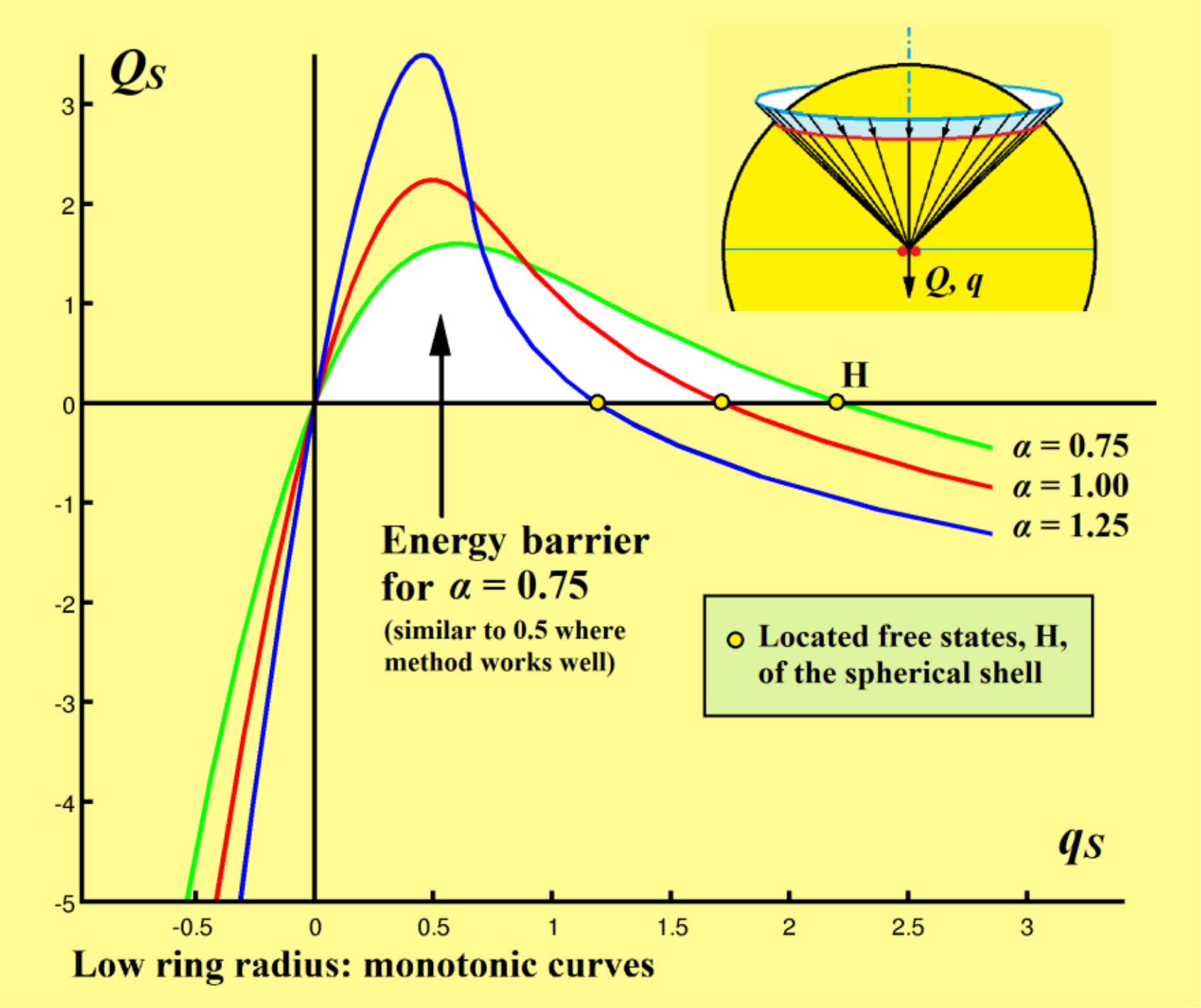

Fig 22. Lateral load deflection curves, $Q_S(q_S)$, under ring-probes of low radius ($\alpha$ = 0.75 to 1.25).

This behaviour becomes modified at medium values of $\alpha$ as illustrated in figure 23 where the curves have developed hysteresis loops between vertical tangencies.

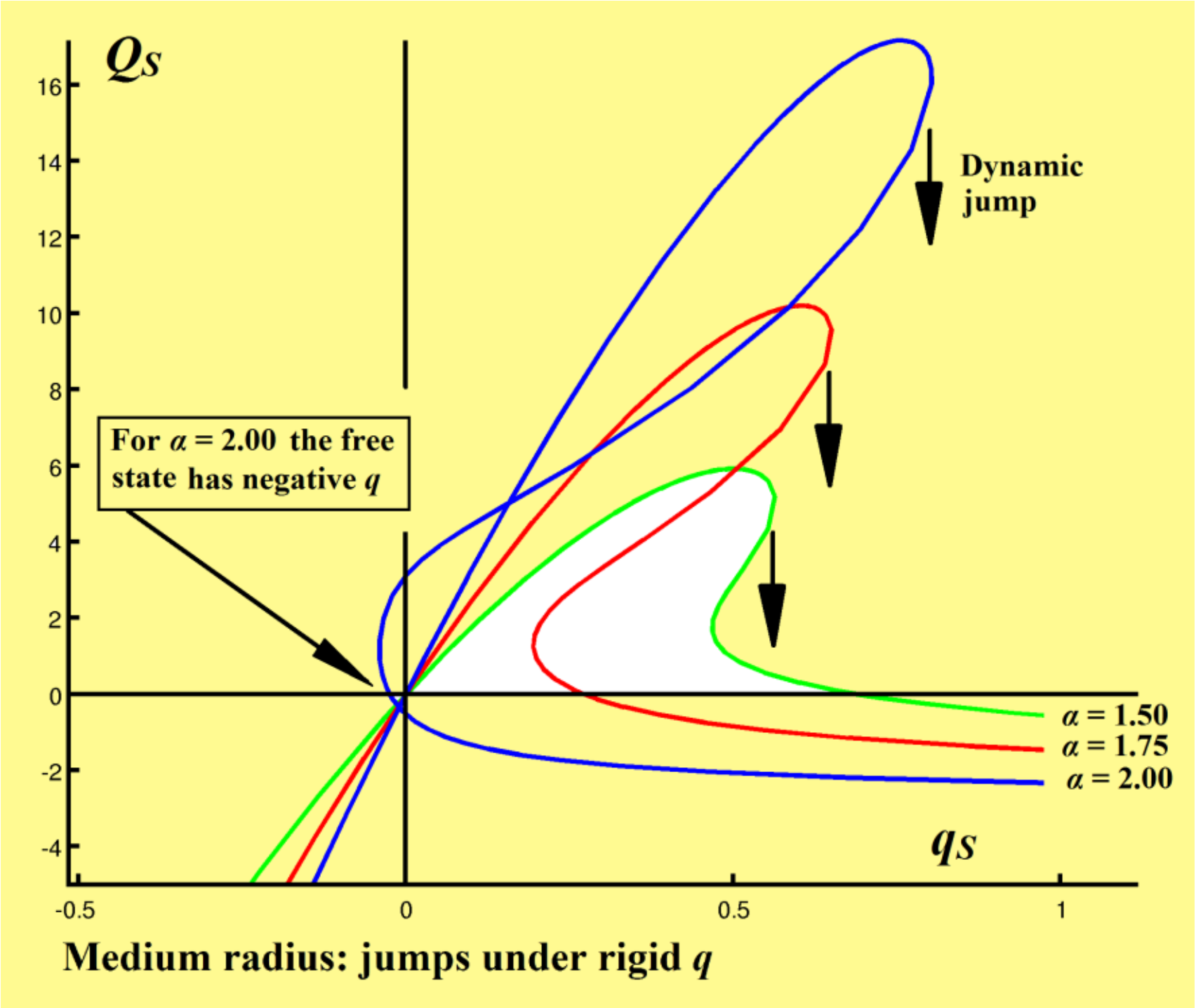

Fig 23. Lateral load deflection curves, $Q_S(q_S)$, under ring-probes of medium radius ($\alpha$ = 1.50 to 2.00).

On loading from the origin under a rigidly controlled displacement, $q_S$, it will now be impossible to follow a curve beyond the first tangency (fold bifurcation) from which a dynamical jump will carry the system to an unknown state (though possibly to the continuation of the same path, especially when close to the cusp generating the hysteresis). This means that the straight forward test with only a single probe will fail to determine the free states of the shell and the required energy barrier. We can also observe that the blue curve continues to a free state of the compression shell with a negative value of $q_S$, a phenomenon that we shall see again in the next figure.

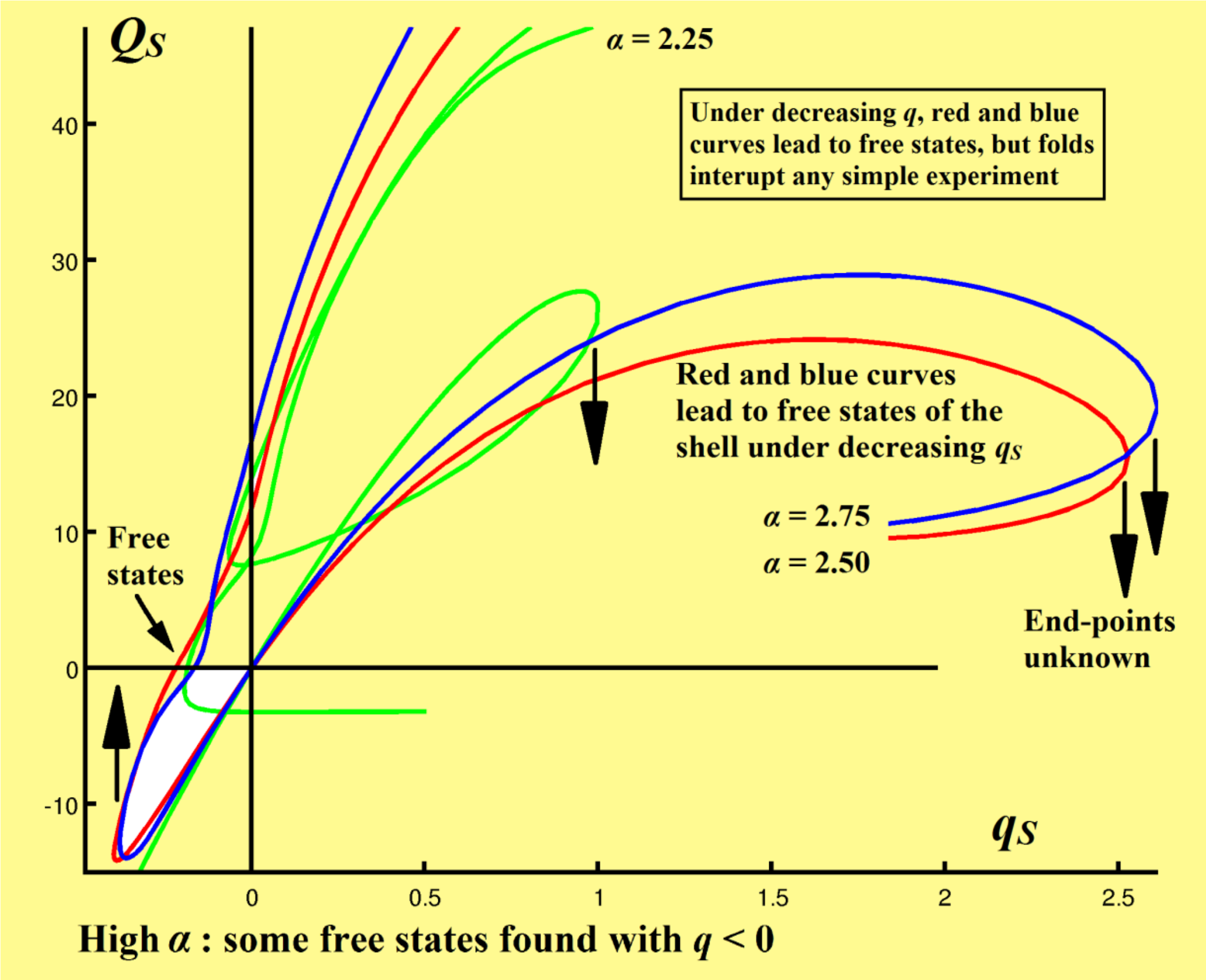

Fig 24. Lateral load deflection curves, $Q_S(q_S)$, under ring-probes of high radius ($\alpha$ = 2.25 to 2.75).

At higher values of the ring radius the observed behaviour becomes even more complex, as illustrated in figure 24. Once again, if we increase $q_S$ from the origin we encounter jumps to unknown states. Interestingly, though, we see that under decreasing $q_S$ the blue and red curves do lead towards free states of the shell with negative $q_S$; but again an experimental test would be interrupted by a jump from a fold bifurcation.

The behaviours that we have just observed remind us that it is just as valid to pull with a probe (that is glued to the shell) as it is to push with a probe. This might work well if a shell (perhaps with an unusual geometry) wanted to buckle outwards at the position of the probe, but as we have just seen it might (as always) fail because of a vertical tangency.

### 6. Possible Problems needing Extra Controls

The first proof of concept in [Thompson, 2015] was provided by Jan Sieber (co-author of the current paper) who analysed the axially compressed cylindrical shell with a side probe as illustrated in figure 25.

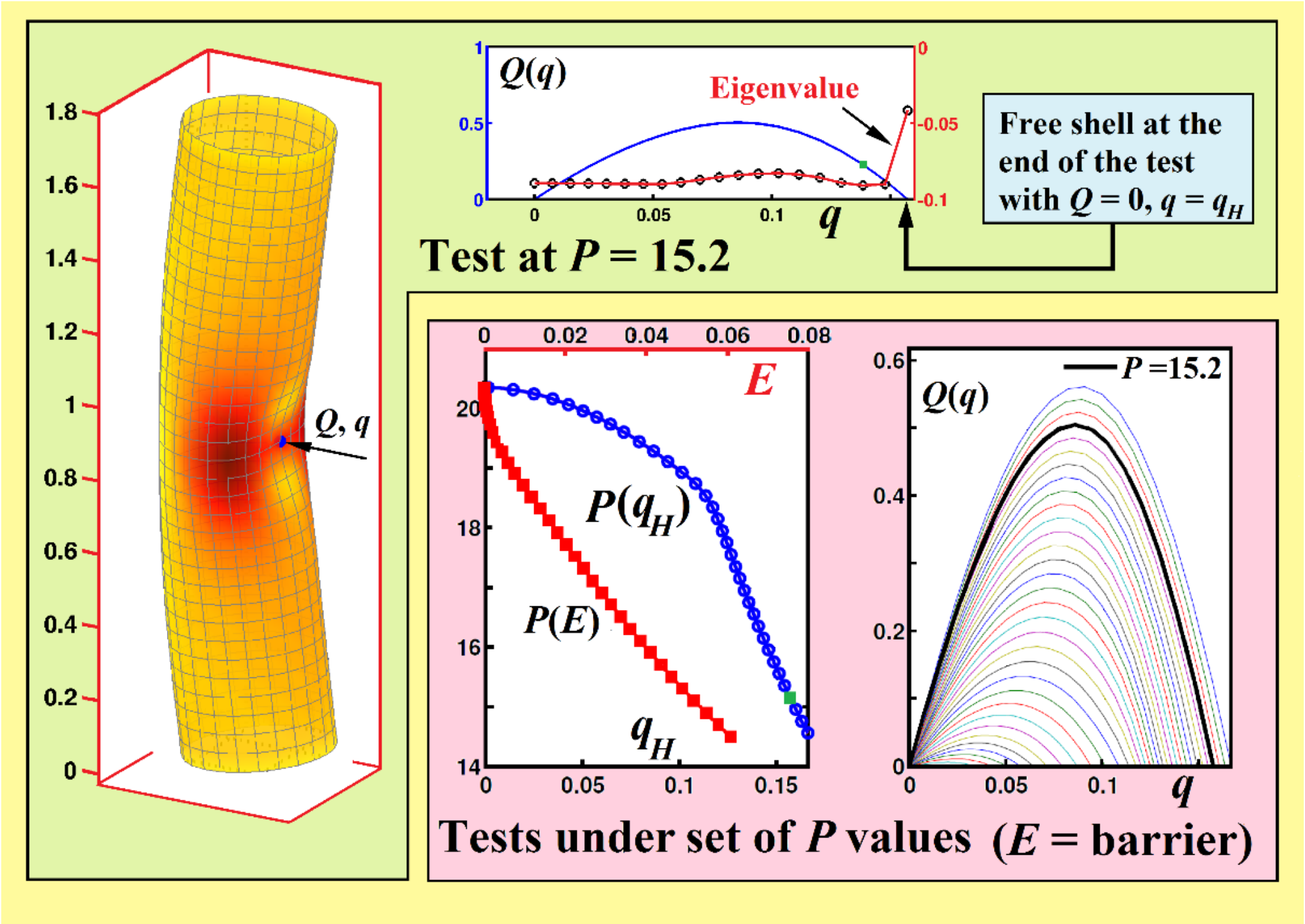

Fig 25. Original proof of concept by Jan Sieber in simulations of a discrete dynamical model of a cylindrical shell [Thompson 2015]. Here the use of a single lateral probe was successful in location the post-buckling curve (H) and providing the shock sensitivity curve $P(E)$, which is exact within the theoretical model.

This was a dynamic analysis, which confirmed that the shell-like link model with 24 particles on each horizontal ring and 35 rings (840 degrees of freedom) remained stable under the imposition of the displacement, $q$, all the way to the free post-buckling saddle, H, at $Q = 0$, $q = q_H$.

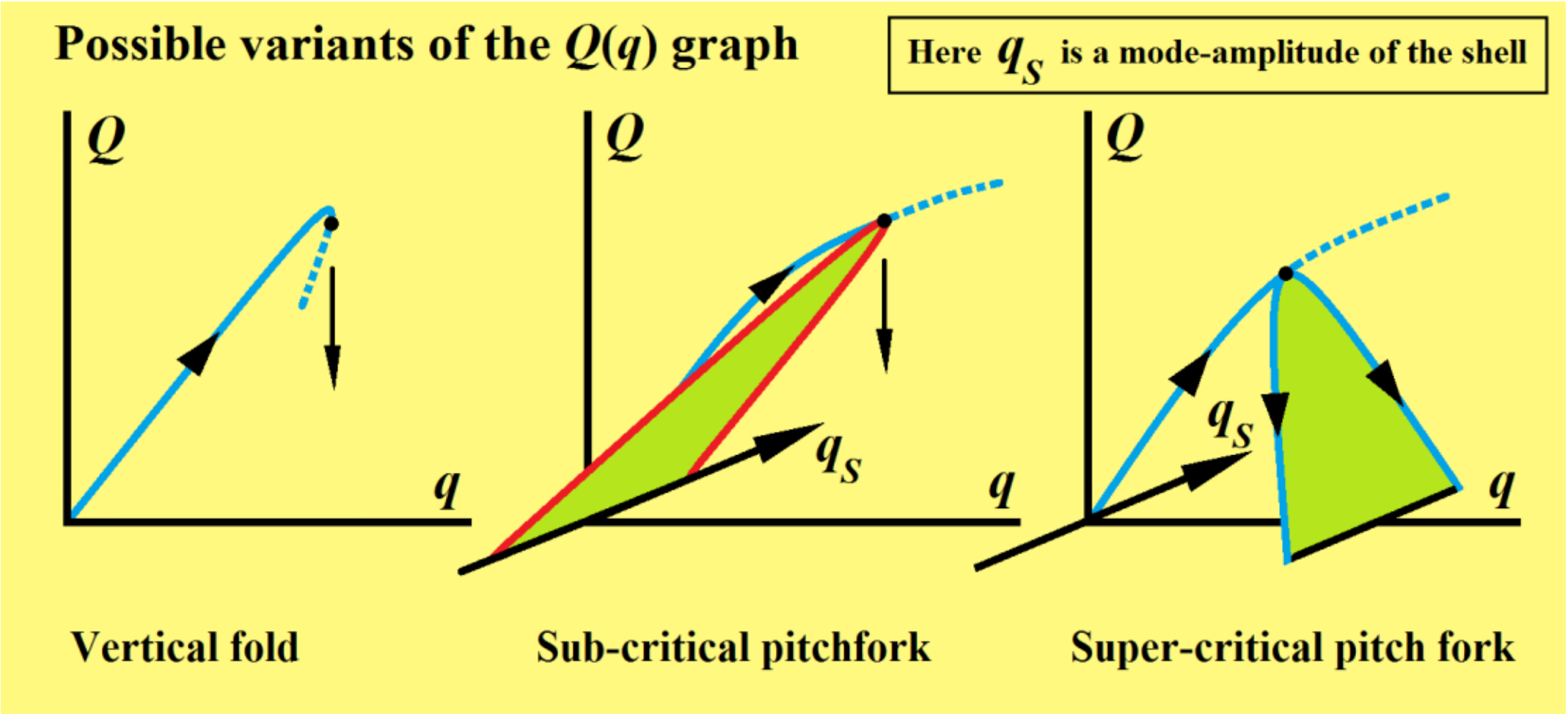

Fig 26. Three possible instability problems that can arise in the lateral $Q(q)$ response under a single rigidly controlled probe.

Possible complications were clearly going to be those illustrated in figure 26 and we shall show shortly how these can (sometimes) be overcome by the introduction of extra controls. Firstly, the determined $Q(q)$ curve might reach an extreme value of $q$, implying a loss of stability and a dynamic jump even under the rigid control of $q$, and we have indeed just seen an example of this in the spherical shell under a ring load. Secondly, the $Q(q)$ curve might exhibit a pitchfork bifurcation which is sub-critical with respect to the controlled $q$: this would also give rise to a dynamic jump as illustrated. Thirdly, the $Q(q)$ curve might exhibit a pitchfork bifurcation which is super-critical with respect to $q$: on encountering this an experimental test could be expected to follow one or other limb of the new bifurcating path.

An example of a pitch-fork bifurcation occurring in the $Q(q)$ plot before it reaches $q_H$ is nicely illustrated in the second study by Jan Sieber [Thompson, 2015] reproduced here in figure 27. Here, for the same model cylinder, but with different parameters, a symmetry breaking pitchfork bifurcation is detected by the vanishing of one of the eigenvalues, $\lambda$, at point C. Whether this is a subcritical or supercritical with respect to $q$ is not known. This distinct bifurcation, with just a single zero eigenvalue, can clearly be stabilised by the introduction of a second rigid probe, B, as was confirmed in the study. This probe was positioned at a point where the eigenvector corresponding to $\lambda$ has a large deflection. It was tuned to provide zero force, while nevertheless stabilising the system.

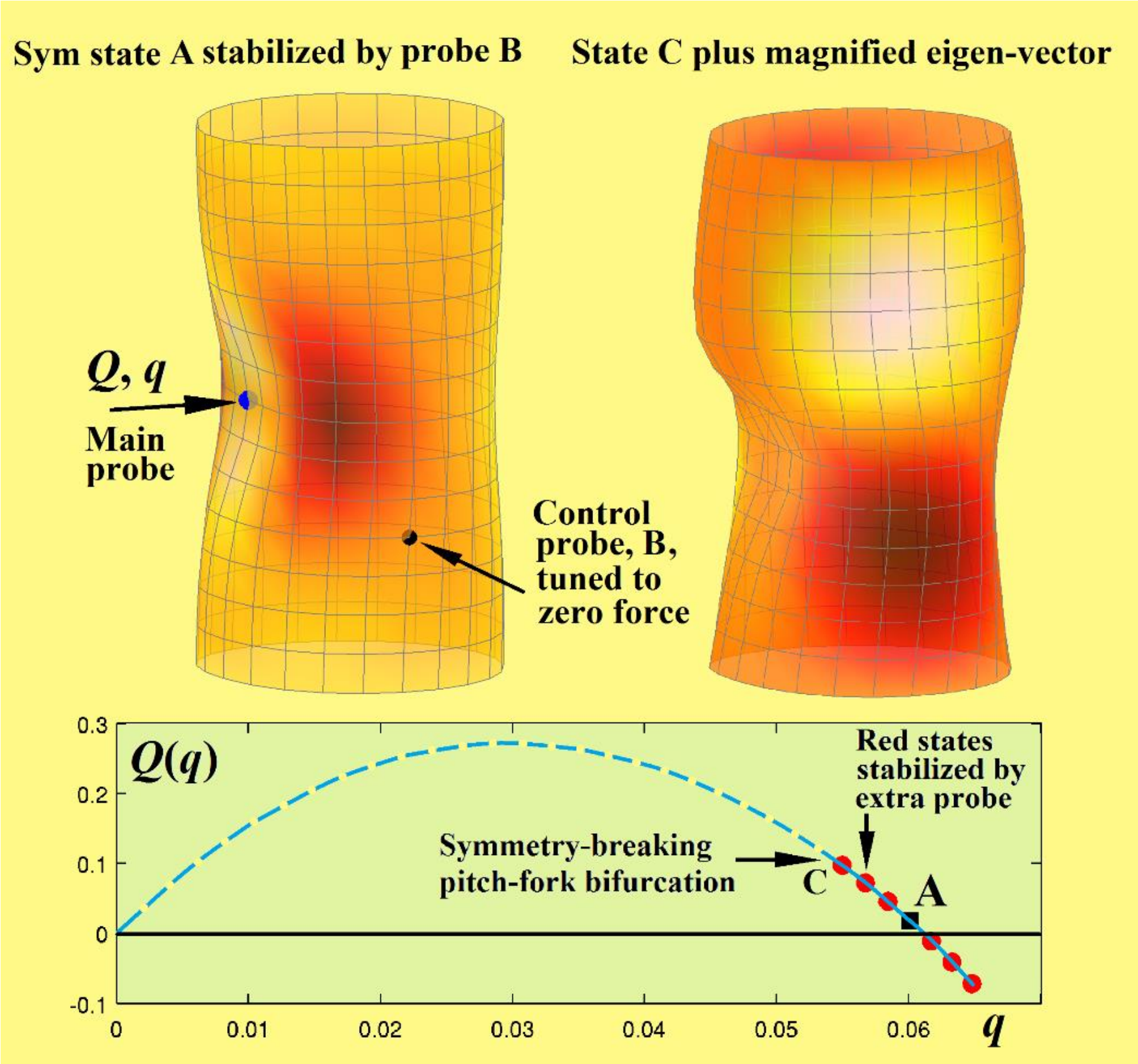

Fig 27. A symmetry-breaking bifurcation found by Jan Sieber in his dynamic shell simulations. Here he demonstrates how the lateral $Q(q)$ curve can be located beyond the bifurcation point, C, by the introduction of a second rigidly controlled point probe, tuned to provide zero force [Thompson, 2015].

The first view of a cylinder in figure 27 shows the symmetric state A, lying just before $q = q_H$ as shown in the lower plot. This can be stabilized beyond the pitch-fork at C by the secondary probe, B. The second view of the cylinder shows the magnified eigen-vector at the pitchfork bifurcation, C. Because the simulation was for the main probe being glued to the shell, it has been possible to stabilize the path not only up to, but beyond the free state H. The interested reader should see Barton & Sieber [2013] for other aspects of exploring bifurcations with noninvasive controls.

## 7. Concluding Remarks

We have reviewed and extended some recent ideas about the non-destructive experimental evaluation of shock-sensitivity in shell-like structures. In particular we have shown how mountain-pass saddles of the total potential energy effectively attract loading path to themselves. We have also demonstrated, by computer simulation, how the process can work for a spherical shell loaded by external pressure, using a single probe at the North Pole. Problems that can arise if a probe is in a sense 'inappropriate' have been explored for this shell using a ring-probe.

One practical aspect of "gluing the probe to the shell" should be mentioned here. We have used this expression simply to imply the use of a probe that can provide either positive or negative values of the lateral force, $Q$, as demanded by the shell, thereby constraining only a single degree of freedom (the lateral displacement). In reality, however, gluing or welding would also offer a constraint against rotation, thereby constraining two more degrees of freedom. This could be useful in, for example, stabilising any symmetry-breaking along the path, but would have to be monitored to keep the constraining moments equal to zero (and, thus, non-invasive).

We can finally observe that the method can be extended easily to other long and thin structures such as, for example, the stretched and twisted rod [Thompson & Champneys 1996, van der Heijden & Thompson 2000, van der Heijden 2001, van der Heijden, *et al* 2002] which has supplied a rich source of elastic buckling phenomena paralleling those of the cylindrical shell [Hunt *et al* 1989, Hunt *et al* 2003, Hunt *et al* 2000].

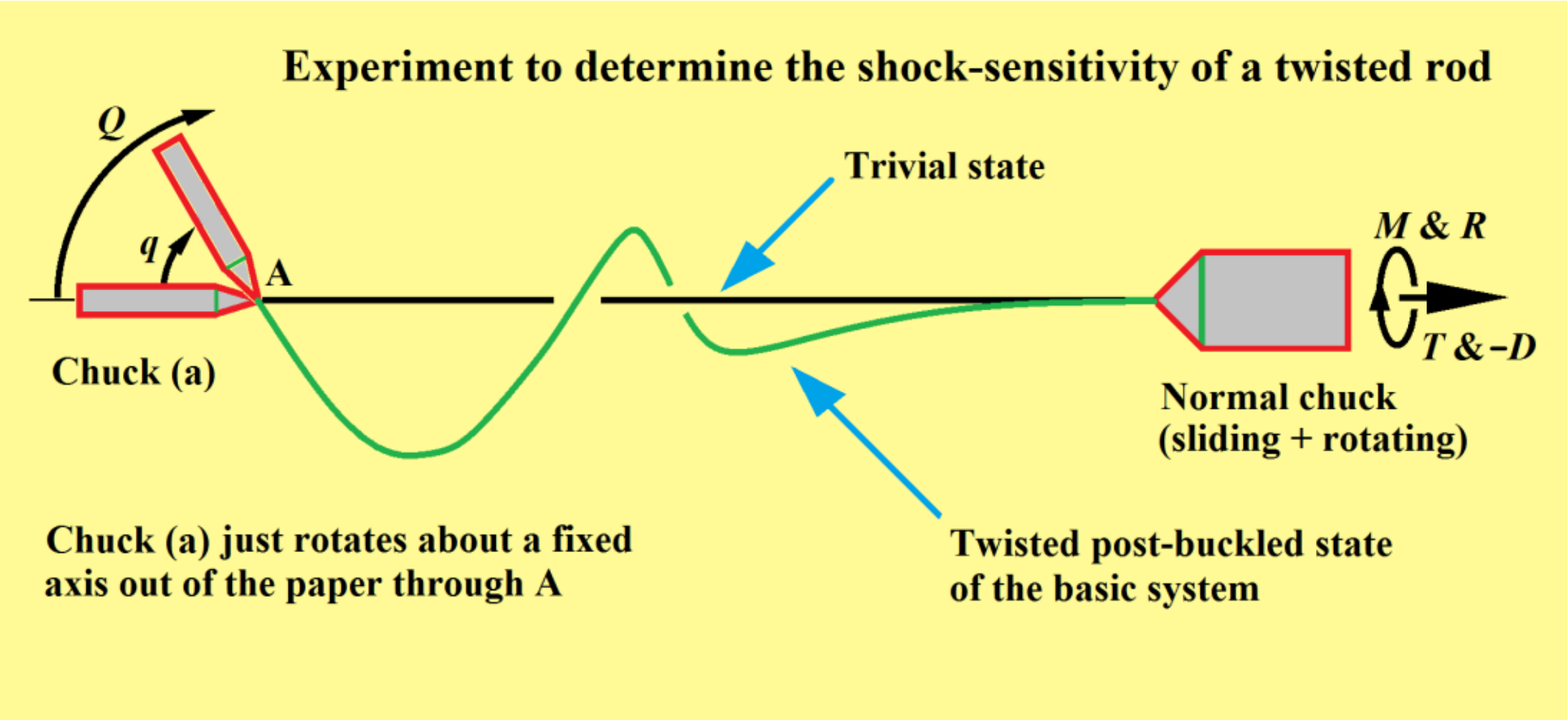

Fig 28. Possible scenario for using a rigidly controlled single probe to assess the shock sensitivity of a stretched and twisted rod.

Here, as an example, we show in figure 28 a rod which is being buckled into a helical form by the sliding and/or rotating of the normal chuck at the right-hand end. The rigid probing is provided by the special chuck (a) at the other end which pivots about an axis out-of-the-paper at A. This pivoting is free for the basic system, which can therefore buckle from its straight trivial configuration into the post-buckled state illustrated. To assess the shock-sensitivity of the trivial state before it buckles, chuck (a) can be rigidly loaded by a rotation, $q$, while the resisting moment, $Q$, of the rod is monitored. In this way we can determine a $Q(q)$ curve, and proceed as for the shells analysed in this paper.

For more pictures and references related to the wider field of shell bucking see the excellent and comprehensive web-page of Bushnell [2015], and for design implications in aeronautics and astronautics [Bushnell, 1981]. The approximate analytical procedures of James Croll are nicely summarised in a recent paper by Godoy, *et al* [2015], and for the probabilistic approach to imperfections we can recommend the new book by Elishakoff [2014]. Other applications may lie in the wrinkling of membranes and thin films, where Takei, et al [2014] make use of a probe in their theoretical analysis.

## Appendix

The coefficients that we used in the spherical shell analysis are shown below in figure 29.

| **Tensor** | $i$ | $j$ | $k$ | $\ell$ | value |
|---|---|---|---|---|---|
| K | 1 | 1 | | | 0.99752 |
| K | 2 | 2 | | | 0.39901 |
| K | 3 | 3 | | | 2.052 |
| K | 1 | 2 | | | 0.24938 |
| K | 1 | 3 | | | −0.14963 |
| K | 2 | 3 | | | −0.59852 |
| C | 1 | 1 | | | 0.21028 |
| C | 2 | 2 | | | 0.011654 |
| C | 3 | 3 | | | 0.0056559 |
| C | 1 | 2 | | | 0.044237 |
| C | 1 | 3 | | | −0.018923 |
| C | 2 | 3 | | | −0.0065165 |
| D | 1 | 1 | | | −0.66667 |
| D | 2 | 2 | | | −0.066667 |
| D | 3 | 3 | | | −0.14286 |
| D | 1 | 2 | | | −0.066667 |
| D | 1 | 3 | | | −0.066667 |
| D | 2 | 3 | | | 0.047619 |

| **Tensor** | $i$ | $j$ | $k$ | $\ell$ | value |
|---|---|---|---|---|---|
| F | 1 | 1 | 1 | | −0.44278 |
| F | 2 | 2 | 2 | | −0.0086886 |
| F | 3 | 3 | 3 | | 0.0089943 |
| F | 1 | 1 | 2 | | −0.048957 |
| F | 1 | 1 | 3 | | −0.025898 |
| F | 2 | 2 | 1 | | −0.020552 |
| F | 2 | 2 | 3 | | 0.0056333 |
| F | 3 | 3 | 1 | | −0.028152 |
| F | 3 | 3 | 2 | | −0.0086887 |
| F | 1 | 2 | 3 | | 0.007691 |
| G | 1 | 1 | 1 | 1 | 0.23729 |
| G | 2 | 2 | 2 | 2 | 0.0020263 |
| G | 3 | 3 | 3 | 3 | 0.0087555 |
| G | 1 | 1 | 1 | 2 | 0.013007 |
| G | 1 | 1 | 1 | 3 | 0.031002 |
| G | 2 | 2 | 2 | 1 | 0.0014202 |
| G | 2 | 2 | 2 | 3 | −0.0012687 |
| G | 3 | 3 | 3 | 1 | 0.0052735 |
| G | 3 | 3 | 3 | 2 | −0.002725 |
| G | 1 | 1 | 2 | 2 | 0.0091745 |
| G | 1 | 1 | 3 | 3 | 0.018165 |
| G | 2 | 2 | 3 | 3 | 0.0019887 |
| G | 1 | 1 | 2 | 3 | −0.0035836 |
| G | 2 | 2 | 1 | 3 | −0.000032484 |
| G | 3 | 3 | 1 | 2 | 0.00023371 |

Fig 29. Numerical coefficients used in the spherical shell analysis